## IMPORTANT NOTE ON THE MATERIAL IN THE THESIS

- The contents of this PhD thesis (beginning August 2010 and ending August 2013) on *Newtonian* drop secondary atomization appear as published articles in the following journals and conference proceedings.

- Further results/papers and publications on non-Newtonian drop atomization are available on request by emailing the author on the following email address.

✉ Please email Varun Kulkarni at varun14kul@gmail.com for any further information or details.

### Journals

3. "*On interdependence of instabilities and drop sizes in bag breakup*", V. Kulkarni ✉, N. Shirdade, N.S. Rodrigues, V. Radhakrishna, P.E. Sojka, June 2023 (in press *Applied Physics Letters*)

2. "*Bag Breakup of Viscous Drops in a Continuous Air Jet.*", V. Kulkarni ✉, P.E. Sojka, ***Physics of Fluids***, Jul 2014, Vol 25(7), pp 072103.

Also please check the arXiv submission for updated and annonated version of all mathematical derivations in the paper on this link. https://arxiv.org/abs/2204.06036

### Peer Reviewed Conference Articles

1. "*Secondary Atomization of Newtonian Liquids in the Bag Breakup Regime: Comparison of Model Predictions to Experimental Data*", V. Kulkarni ✉, D.R. Guildenbecher, P.E. Sojka, ICLASS 2012, 12[th] International Conference on Liquid Atomization and Spray Systems-**ILASS**, Sep 2-6, 2012, Heidelberg, Germany.

### Conference presentations

3. "*Subtle interplay of rim and bag instabilities in drop atomization*", **V. Kulkarni**, N. Shirdade, V. Radhakrishna, 75[th] Annual Division of Fluid Dynamics Meeting (**APS-DFD**) Nov 20-22, 2022, Indianapolis, IN.

2. "*Fragmentation dynamics in the droplet bag breakup regime*", **V. Kulkarni**, P.E. Sojka, 67[th] Annual Division of Fluid Dynamics Meeting (**APS-DFD**), Nov 23-25, 2014, San Francisco, CA.

1. "*Viscous Bag Breakup*", **V. Kulkarni**, D. Guildenbecher, S. Firehammer, P.E. Sojka, 65[th] Annual Division of Fluid Dynamics Meeting (**APS-DFD**), Nov 18-20, 2012, San Diego, CA.

Varun Kulkarni

AN ANALYTICAL AND EXPERIMENTAL STUDY OF SECONDARY
ATOMIZATION FOR VIBRATIONAL AND BAG BREAKUP MODES

A Dissertation

Submitted to the Faculty

of

Purdue University

by

Varun Kulkarni

In Partial Fulfillment of the

Requirements for the Degree

of

Doctor of Philosophy

August 2013

Purdue University

West Lafayette, Indiana



To my parents and brother Vikrant for their support and encouragement.



# ACKNOWLEDGEMENTS

The PhD experience has been extremely satisfying and without the contribution of others it would not have been possible. I wish to acknowledge them here. Foremost, I would like to thank my advisor, Prof. Paul E. Sojka, for giving me the opportunity to pursue a PhD at Purdue University. I feel a deep sense of gratitude towards him and would like to especially thank him for sharing his office with me. His calming presence has been instrumental in ensuring that I never felt the pressures of graduate student life and contributed to me working efficiently. I feel honored to have been associated with a mentor like him. I would like to also thank Prof. S. Heister, Prof. C. Wassgren for serving on my committee, Prof. S. Wereley for allowing me to collaborate with him and Prof. Anil K. Bajaj in his capacity as the head of the School of Mechanical Engineering.

I am grateful to my lab members, Dr. Dan Guildenbecher, Dr. Ariel Muliadi and Dr. Alexis Déchelette for offering me valuable advice whenever needed. I greatly thank Rohan for his support in accomplishing some of the experimental part of this work. It was very kind of him to make himself available at all times. The research carried out with Avanish has been extremely rewarding and has helped me expand my repertoire beyond jet and droplet dynamics. I greatly appreciate his cooperation and efforts.

The graduate office staff at the ME department has been extremely helpful. I thank Prof. Anderson, Cathy Elwell, Gail Biberstein and Julayne Moser for their valuable administrative support. I also extend my gratitude to Charlotte Bell for expediting the paper work which would have otherwise taken much longer.



Research and coffee have always had a deep connection. In Kamesh and Niranjan I found two able mates who made visits to the coffee shop pleasurable. The invigorating discussions about scientific research and its various facets shall remain with me as an enduring memory of the time spent as a graduate student at Purdue.

Outside my work environment, I would like to thank my roommate Pranav for his enjoyable company. It was good to have someone as forward looking and enthusiastic like him around. I am indeed grateful to Rajeev and Vivek for their friendship, which has only grown stronger over the years. Conversations with them have provided the much needed respite from a hard day's work at the office. I thank Nitin and Raviraj for providing me accommodation at crucial junctures of my stay at West Lafayette, and for lightening the atmosphere during some of the intense moments.

There have been many others both inside and outside the university who have in some small way or the other contributed to this work. The confines of this space do not permit me to mention them individually. I place on record my sincere thanks to all of them.

Last but not the least, I am grateful to my family: mom, dad, brother (Vikrant) and sister-in-law (Neha) for providing an intellectually congenial atmosphere at home. They have been pivotal in instilling in me values of sincerity, hard work and unrelenting curiosity, which have held me in good stead in my life so far. I hope it continues in the years to come.



TABLE OF CONTENTS















LIST OF TABLES





LIST OF FIGURES













## NOMENCLATURE

| | |
|---|---|
| $c$ | Speed of sound [$m/s$] |
| $C_D$ | Drag Coefficient |
| $\overline{C_D}$ | Average drag coefficient |
| $D_{10}$ | Arithmetic mean diameter [$m$] |
| $D_{30}$ | Volume mean diameter [$m$] |
| $D_{32}$ | Sauter mean diameter [$m$] |
| $d_0$ | Initial drop diameter [$m$] |
| $d_{cro}$ | Cross−stream drop length [$m$] |
| $d_{str}$ | Streamwise drop length [$m$] |
| $f_0(D)$ | Number pdf [$1/m$] |
| $F_D$ | Drag Force [$N$] |
| g | Gravitational Acceleration [$m/s^2$] |
| $h(t)$ | Rim thickness [$m$] |
| $H(t)$ | Non-dimensional rim thickness |
| $\boldsymbol{I}$ | Identity Tensor |
| $k_r$ | Stiffness constant [$N/m$] |
| $p$ | Thermodynamic Pressure |
| $p_a$ | Air pressure field around the drop [$Pa$] |
| $p_l$ | Liquid pressure field inside the drop [$Pa$] |
| $r_c$ | Radius of curvature at the bag tip [$m$] |
| $R(t)$ | Radial extent of the bag [$m$] |
| $t$ | Dimensional time [$s$] |
| $T$ | Non-Dimensional time |
| $\boldsymbol{T}$ | Stress tensor [$N/m^2$] |
| $T_{ini}$ | Non-Dimensional initiation time |
| $T_{rr}(l)$ | Liquid normal stress components [$N/m^2$] |
| $T_{rr}(g)$ | Gas (here, air) normal stress components [$N/m^2$] |
| $u_r(r,t)$ | Radial velocity field inside the drop [$m/s$] |
| $u_{r_c}$ | Liquid velocity field inside the bag [m/s] |
| $U_r$ | Air velocity in the radial direction [$m/s$] |
| $U_{rel}$ | Mean relative air velocity [$m/s$] |
| $U_y$ | Air velocity in the streamwise direction [$m/s$] |
| $Y$ | Ratio of the cross-stream diameter to the spherical diameter |
| $\alpha(t)$ | Maximum bag extent [$m$] |
| $\beta(t)$ | Non-dimensional maximum bag extent |
| $\gamma$ | Acceleration of the liquid interface [$m/s^2$] |
| $\mu_l$ | Liquid dynamic viscosity [$Pa-s$] |
| $\mu_g$ | Gas dynamic viscosity [$Pa-s$] |
| $\phi(t)$ | Non-dimensional radial extent of the bag |
| $\phi_{max}$ | Value of $\phi(t)$ before bag formation |
| $\rho_l$ | Liquid density [$kg/m^3$] |
| $\rho_g$ | Gas density [$kg/m^3$] |
| $\sigma$ | Surface tension of the liquid/gas interface [$N/m$] |
| $\kappa$ | Curvature of the interface [$1/m$] |
| $\tau$ | Characteristic time scale for the atomization process [$s$] |
| $\tau_c$ | Characteristic time scale for Plateau–Rayleigh and Rayleigh–Taylor instabilities [$s$] |

Dimensionless Groups

| | |
|---|---|
| $\varepsilon$ | Ratio of liquid to gas density |
| $Bo$ | Bond Number |
| $La$ | Laplace Number |
| $N$ | Ratio of liquid to gas viscosity |
| $Oh$ | Ohnesorge Number |
| $We$ | Weber Number |
| $We_c$ | Critical Weber Number |
| $We_{c\,Oh\to 0}$ | Critical Weber Number at $Oh \to 0$ |
| $We_{32}$ | Weber Number based on Sauter mean diameter, $D_{32}$ |

The overdots represent derivatives *w.r.t* non-dimensional time, $T$



# LIST OF ABBREVIATIONS

| | |
|---|---|
| DNS | Direct Numeric Simulation |
| LDA | Laser Doppler Anemometry |
| MMD | Mass mean diameter |
| MEF | Maximum Entropy Formulation |
| MPS | Moving-Particle Semi-implicit |
| PDA | Phase Doppler Anemometry |
| PIV | Particle Image Velocimetry |
| PMT | Photo Multiplier Tube |
| PTFE | Poly-tetra-fluoro-ethylene |
| PVC | Poly-vinyl-chloride |
| SNR | Signal-to-Noise Ratio |
| UV | Ultra Violet |
| VOF | Volume of Fluid |



ABSTRACT


Kulkarni, Varun Ph.D., Purdue University, August 2013. An Analytical and Experimental Study of Secondary Atomization for Vibrational and Bag Breakup Modes. Major Professor: Dr. Paul Sojka, School of Mechanical Engineering.

Bag breakup of drops has been a subject of interest for almost over a century. Several issues such as theoretical estimation of the regime boundary marking the onset of such breakup, bag growth rates, drop size distribution, and the effect of Weber number, $We$, and Ohnesorge number, $Oh$, on these quantities remain unaddressed even in works as recent as those of Zhao *et al*. (2010) and Cao *et al*. (2007).

The current study aims to clarify aspects of the atomization process through experiments and theory. We examine bag breakup of a single drop of various inviscid and low viscosity fluids as it deforms in the presence of a continuous horizontal air jet. The $We$ boundary at which bag breakup begins is theoretically determined and the expression obtained, $We = 12\left(1 + \frac{2}{3}Oh^2\right)$, is found to match well with experimental data of Hsiang and Faeth (1995) and Brodkey (1967). An exponential growth in the radial extent of the deformed drop and the streamline dimension of the bag is predicted by the theoretical model and confirmed by experimental findings. These quantities are observed to strongly depend on $We$. However, their dependence on $Oh$ is weak for the range of $Oh$ considered in this study. Subsequent to drop deformation, bag formation and expansion is the bursting process. This is marked by the disintegration of the bag owing to instability of the Rayleigh–Taylor type, followed by collapse of the liquid rim bounding this bag by Plateau–Rayleigh instability. The sizes of the drops thus produced are measured using Phase Doppler Anemometry (PDA) which is in contrast to shadowgraphs used in earlier




studies of Chou *et al.* (1998), Zhao *et al.* (2002). A discernible shift in the peak of the drop size distribution for viscous drops is seen which indicates a preponderance of drops of higher diameters vis–à–vis fragment size distribution for inviscid drops. Furthermore, an estimate of the Sauter mean diameter ($D_{32}$) is presented which is somewhat lower than the predictions of Dai and Faeth (2001).



CHAPTER 1.  INTRODUCTION

1.1     Background

The process of atomization is a complex one wherein bulk liquid undergoes a transformation that converts it into small masses (drops), it find uses in a variety of applications. We see occurrences of atomization ranging from everyday life experiences, such as when we open the water faucet at our homes, to more intricate phenomena such as burning of fuel in the combustor of rocket engines. Such a collection of drops in the presence of a gaseous medium constitute what is commonly known as a spray.

Sprays are important in an immense range of applications. Some examples are tablet coatings in the pharmaceutical industry, spray drying of food products, spray quenching during metal manufacturing, spray etching of chemical agents on electronic circuit boards, sprinklers used in firefighting equipment, coatings applied during paper manufacturing, plus painting of automobiles, furniture and appliances.

Atomization leads to a myriad of fragment sizes which are rarely uniform. Thus, controlling drop sizes is vital, difficult and a common priority when achieving a desired objective. Depending on the application, sometimes we need drops of larger size (agricultural sprays where we do not want to very small drop sizes or else spray drift becomes a problem) or smaller drop diameters (respirable/inhalable sprays for medical treatments).  In contrast, in most engine combustion applications we seek a range of drop sizes so that we get high combustion efficiency, good ignition, and a flame that is spread throughout the combustion chamber.   Regardless, the ability to accurately control drop



sizes depends in part on being able to predict said sizes. Such predictions are central to this thesis research.

The atomization process is actually two-fold, beginning with the initial bulk liquid formed into drops in a process called primary atomization. The final stage is fragmentation of drops into smaller drops; this is called secondary atomization. Secondary atomization is more pronounced when the drops either have higher relative velocities compared to the surrounding gas or are large in size themselves.

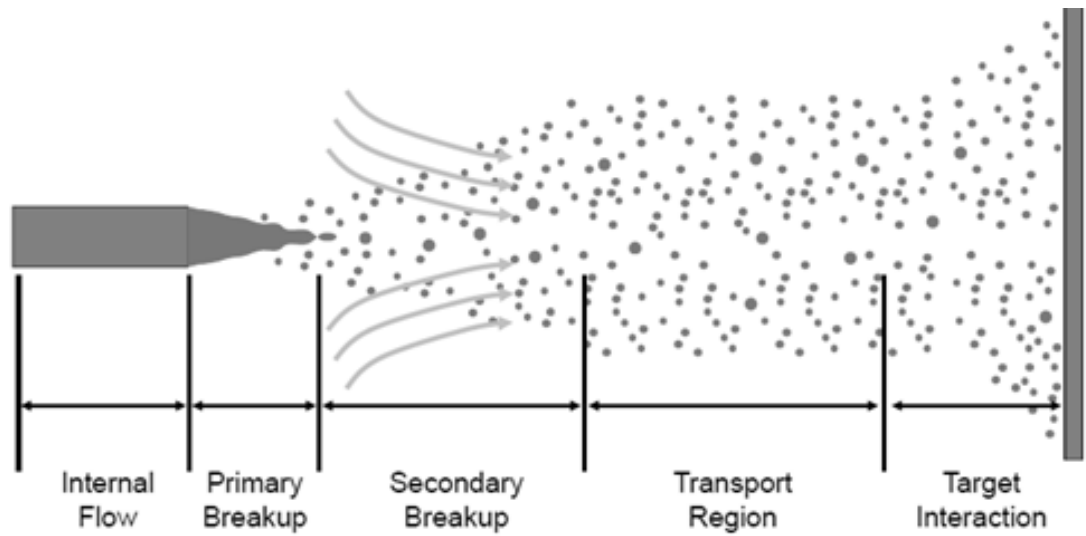

Figure 1.1: The various stages of breakup of bulk liquid as it exits the atomizer and moves towards the target (Guildenbecher *et al.*, 2009).

Clearly seen in Fig. 1.1 are zones of primary and secondary breakup. The curved arrows show entrainment, which draws the surrounding gas inside the spray sheath thereby creating a region where the liquid is dispersed throughout the gaseous medium. Even if the original drops are moving at the same speed as the surrounding gas, they will undergo deceleration due to the surrounding gas slowing down within the spray. The end result is further breakup leading to a distribution of drop sizes.



From the point of view of fluid mechanics, these drops can either be Newtonian or non−Newtonian in nature. Newtonian fluids are those where the relation between fluid stress and the rate of strain and can be expressed by the simple constitutive relation,

$$\boldsymbol{T} = -p\mathbf{I} + \mu_l \left[ \boldsymbol{\nabla} u + \left( \boldsymbol{\nabla} u \right)^{\mathrm{T}} \right].$$ 
(1.1)

In (1.1) $p$ is the thermodynamic pressure, $u$ the velocity field, $\mu_l$ the dynamic viscosity (a material property) and $\boldsymbol{T}$ the stress tensor. Fluids with $\mu_l = 0$ are termed inviscid, and those with $\mu_l \neq 0$ are known as viscous fluids.

## 1.2 Secondary Atomization

Disruptive aerodynamic forces play a key role in secondary atomization. Their presence is required for drop breakup to occur. In addition, their relative magnitude is an important consideration when predicting the manner in which a drop will break up. An important dimensionless number in this regard is the Weber number,

$$We = \frac{\rho_g U_{rel}^2 d_0}{\sigma}.$$ 
(1.2)

where $\rho_g$ is the density of the gas surrounding the drop, $U_{rel}$ is the initial relative velocity between the gas and the drop, $d_0$ is the initial drop diameter, and $\sigma$ is the interfacial (surface) tension. In the context of Newtonian drops we use the Ohnosorge number to account for drop viscosity,

$$Oh = \frac{\mu_l}{\sqrt{\rho_l \sigma d_0}}.$$ 
(1.3)

where $\mu_l$ is the dynamic viscosity of the drop and $\rho_l$ its density. While $We$ is a measure of the ratio of aerodynamic to restoring forces, $Oh$ indicates the ratio of viscous to surface tension forces. Higher $Oh$ implies more viscous drops.

Many researchers, beginning with Lenard (1904), and continuing with Hinze (1946), Ranger and Nicholls (1969), Krzeczkowski (1980), Pilch and Erdman (1987), Wierzba



and Takayama (1988), Hsiang and Faeth (1992, 1995), Liu and Reitz (1997) and Chou and Faeth (1998), have studied Newtonian drop secondary breakup, identified the various regimes in which breakup occurs, and have listed transitions between one regime and its neighbors based on We. Five common modes of breakup have been agreed upon for low *Oh* conditions,

1. Vibrational ($0 < We < \sim 11$)
2. Bag ($\sim 11 < We < \sim 35$)
3. Bag and Stamen ($\sim 35 < We < \sim 80$)
4. Sheet Thinning ($\sim 80 < We < \sim 350$)
5. Catastrophic ($We > \sim 350$)

More details on these dimensionaless groups and breakup regimes are provided in Chapter 2.

As will be demonstrated in Chapter 2, the ability to predict drop sizes due to secondary breakup is in its infancy. The objective of this thesis is to resolve issues concerning one aspect of that breakup, the bag breakup regime of Newtonian drop secondary atomization. Several associated parameters such as initiation time, bag growth, rim size, bag thickness, and mean fragment size are quantified theoretically and experimentally. Next, an analytical model to describe the first stage of secondary breakup, drop deformation, is formulated. An analytical model for bag growth is next derived, and predictions using it are compared to experimental data. A discussion of similarities and differences follows, along with a physical explanation for the experimentally observed behavior developed using the analytical model predictions. A summary of the findings is provided in the last chapter, along with the main conclusions drawn from this investigation. The document closes with an outline of potential future work.



## 1.3    <u>Layout of Thesis</u>

The chapters in this thesis have been organized in the following fashion. Chapter 1 gives an introduction to atomization in general, and some preliminaries on secondary atomization. Chapter 2 presents a comprehensive review of the literature. This is followed by Chapter 3, which presents details of the experimental apparatus and the measurement techniques, along with their accuracy, that were used in this study. Theoretical modeling of the bag breakup process is discussed in Chapter 4. The experimental results and their comparison to analytical model predictions are given in Chapter 5. Chapter 6 summarizes the conclusions of this work and indicates areas where future work is possible.



CHAPTER 2.  LITERATURE SURVEY

2.1     <u>Overview</u>

The study of Newtonian drop secondary atomization dates back at least to Lenard's work in 1904. The field is rich in phenomena, so numerous studies have followed. The literature has been reviewed by several groups, the latest being Guildenbecher *et al*. (2009). Little new material has been added since that review, except for a recent study by Theofanus (2011) on viscoelastic fluids, which will be discussed because it provides some of the latest thoughts on secondary atomization (although not completely relevant to our study, since our drops are Newtonian).

Drop deformation and breakup processes can be achieved under laboratory conditions by one of the following methods:
- Drop Tower
- Shock Tube
- Continuous Jet

The methods vary in the type of aerodynamic loading the drop is subject to while it moves.
In a drop tower the drops fall due to gravity. The change in velocity is usually gradual, in contrast to the other two methods where it is stepwise (shock tube) or nearly so (continuous jet). Natural phenomena such as raindrops are studied using drop towers. Villermaux and Bossa (2009) recently used this method to study raindrop fragmentation.



In the continuous jet method the drops enter a jet of high speed air. Care must be taken to ensure that the drop moves through a region of constant, or at least nearly constant, velocity thereby subjecting it to the same kind of loading as in a shock tube. A major advantage of continuous jets is that they allow relatively rapid measurement of fragment size distributions because a drop train can be directed into the jet and fragments can be sampled automatically. Relevant studies which used continuous jets include those of Liu *et al*. (1993), Liu and Reitz (1993), Arcoumanis *et al*. (1996), Hwang *et al*. (1996), Liu and Reitz (1997), Lee and Reitz (2001), Park *et al*. (2006), and Cao *et al*. (2007).

A shock tube contains two sections divided by a diaphragm. Pressurized gas is released from the driver section into the driven section when the diaphragm bursts (a shock wave develops and travels down the tube). Droplets are inserted into the driven section and there is breakup observed. This method has been used by Hinze (1955), Ranger and Nicholls (1969), Gelfand *et al*. (1973), Wierzba and Takayama (1988), Hsiang and Faeth (1992, 1993, 1995), Chou *et al*. (1997), Chou and Faeth (1998), Joseph *et al*. (1999), Igra and Takayama (2001), Dai and Faeth (2001), Joseph *et al*. (2002), Igra *et al*. (2002), and Theofanous *et al*. (2004).

In this thesis the continuous jet approach is used as it is easier to put into practice in the laboratory. It also allows for quick measurements of fragment sizes. As will be seen in Chapter 3, continuous jet results are comparable to shock tube results since the aerodynamic loading is similar to that developed by a shock tube.

## 2.2    Dimensional Analysis

Drop fragmentation is a complex phenomenon with numerous physical mechanisms contributing to the final state. Therefore, it is important to identify the factors affecting the process, the factors that delineate between various behaviors, and the determining parameters to facilitate comparison between these behaviors. The obvious choice for Newtonian drops would be air (or other surrounding gas) density, drop dynamic viscosity,



surface tension, relative velocity (between the drop and its surroundings), and the diameter of the drop. Additional choices could include liquid density and air viscosity. Using dimensional analysis we can establish the following non-dimensional groups:

$$Re = \frac{\rho_g U_{rel} d_0}{\mu_g}. \tag{2.1}$$

$$We = \frac{\rho_g^2 U_{rel}^2 d_0}{\sigma}. \tag{2.2}$$

$$Oh = \frac{\mu_l}{\sqrt{\sigma \rho_l d_0}}. \tag{2.3}$$

In these equations, $\rho_g$ is the gas density typically air, $U_{rel}$ is the initial relative velocity between the drop and its surroundings, $d_0$ is the drop initial diameter, $\mu_l$ is the liquid dynamic viscosity, and $\sigma$ the surface tension. $Re$ is, of course, the Reynolds number, while $We$ is the Weber number. Other non-dimensional groups that can be formed by a combination of the above are the Laplace number, $La$, and Ohnesorge number, $Oh$ which are defined below.

$$La = \frac{1}{Oh^2}. \tag{2.4}$$

$Oh$ and $We$ are the most commonly used. $Oh$, gives a measure of the viscous resistance offered by the drop to the external deformation. The higher the value of $Oh$, the more time it typically takes for the drop to deform. $We$, on the other hand, gives a measure of the aerodynamic disruptive forces relative to the restoring force due to surface tension. As $We$ increases, the effects of aerodynamics forces become ever more important. In addition to the above, the Bond or Eötvös number is usually defined to account for the effect of gravity. This is given below:

$$Bo = \frac{\rho_l g d_0^2}{\sigma}. \tag{2.5}$$



At larger *Bo,* the effect of gravity is more pronounced. In (2.5) 'g' can be replaced by an acceleration term in cases where the drop does not fall down due to gravity. Table 2.1 contains a comprehensive list of these numbers.

The importance of compressibility can be ascertained by computing *Ma.* In situations where the surrounding environment density and viscosity approach values for the drops ε and *N* may play a role (i.e., diesel engines, rocket motor chambers, and high pressure-ratio gas turbines). Additional effects, such as turbulence within the two fluids or unsteady ambient flow, may also lead to destabilization of drops. Impurities and particulates within drops can serves as nucleation sites for breakup, thereby enhancing secondary atomization. Gelfand (1996) and Clift *et al*. (1978) have discussed the impact of such factors. They are believed not to be important to this study.

Table 2.l: Dimensionless groups for Newtonian drop secondary atomization.

| | |
|---|---|
| *We* | $\dfrac{\rho_g U_{rel}^2 d_0^2}{\sigma}$ |
| *Oh* | $\dfrac{\mu_l}{\sqrt{\rho_l \sigma d_0}}$ |
| *Re* | $\dfrac{\rho_g U_{rel} d_0}{\mu_l}$ |
| *Ma* | $\dfrac{U_{rel}}{c}$ |
| *ε* | $\dfrac{\rho_l}{\rho_g}$ |
| *N* | $\dfrac{\mu_l}{\mu_g}$ |



Similar to the non-dimensional groups listed in Tables 2.1 and 2.2, it is useful to define a non dimensional time when describing the scale on which breakup occurs. A non-dimensional time based on drop transport, as introduced by Ranger and Nicholls (1969), is commonly used.

$$T = \frac{U_{rel}}{\sqrt{\varepsilon}\, d_0} t.$$  (2.6)

Here, $T$ is the dimensionless time, $t$ is the dimensional time, $V_0$ is the initial relative velocity between drop and ambient, $\varepsilon$ is the drop-to-ambient density ratio, and $d_0$ is the initial spherical diameter.

While commonly used, equation (2.6) is not appropriate to describe all temporal phenomena in secondary atomization. For example, Shraiber *et al*. (1996) have suggested non-dimensionalizing by the drop oscillation period, while Faeth *et al*. (1995) have recommended using a viscous timescale for drops at high *Oh*.

2.3    <u>Breakup Modes</u>

Increasing the relative velocity between the drop and its surroundings leads to an increase in the disruptive forces which lead to breakup. In the following, a discussion is presented on how this occurs for Newtonian and non-Newtonian drops. This is followed by a detailed discussion of regime boundaries for all cases.

2.3.1    <u>Newtonian Liquids</u>

The breakup of Newtonian liquids has been the subject of study for a long time. Pilch and Erdman (1986) summarize the findings of earlier researchers and the modes of breakup observed



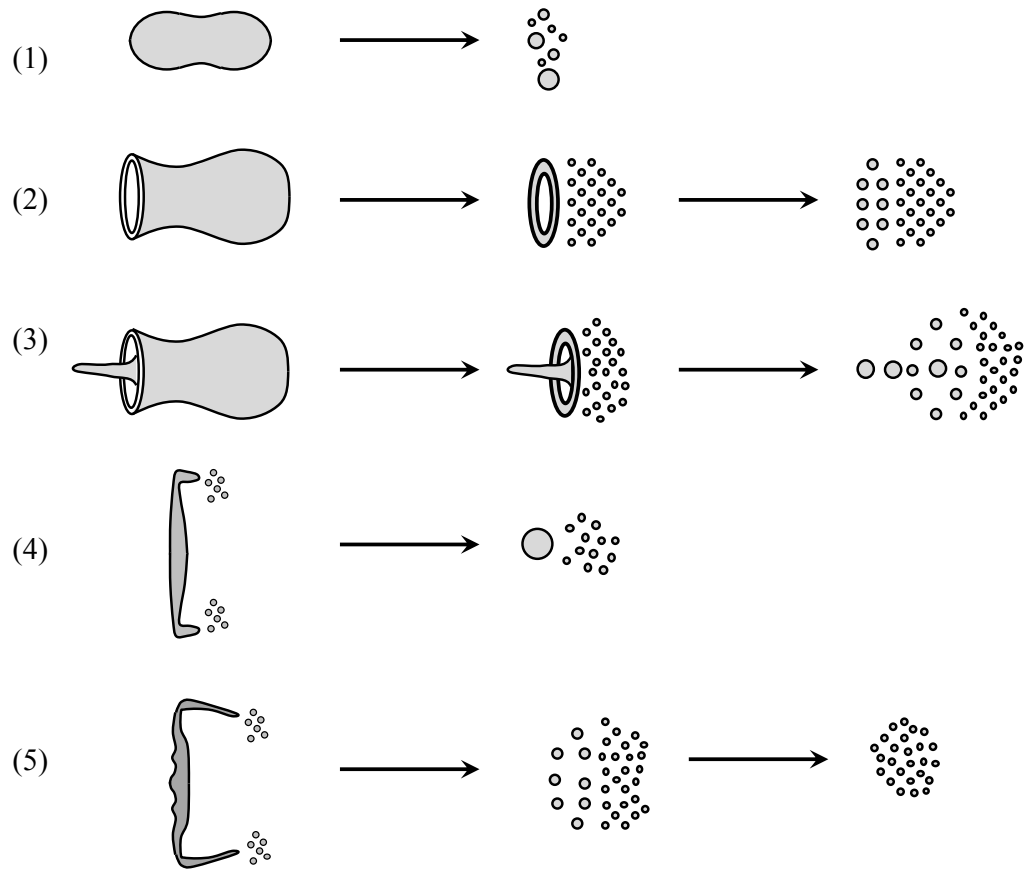

Figure 2.1: Five regimes of breakup for a low viscosity (inviscid) Newtonian drop in the presence of a horizontally flowing air stream, Pilch and Erdman (1987): (1) vibrational ($We \leq \sim 11$) (2) bag (focus of present study) ($\sim 11 \leq We \leq \sim 35$) (3) bag and stamen, or multi-mode ($\sim 35 \leq We \leq \sim 80$) (4) sheet thinning, or shear stripping ($\sim 80 \leq We \leq \sim 350$) (5) catastrophic ($We \leq \sim 350$).



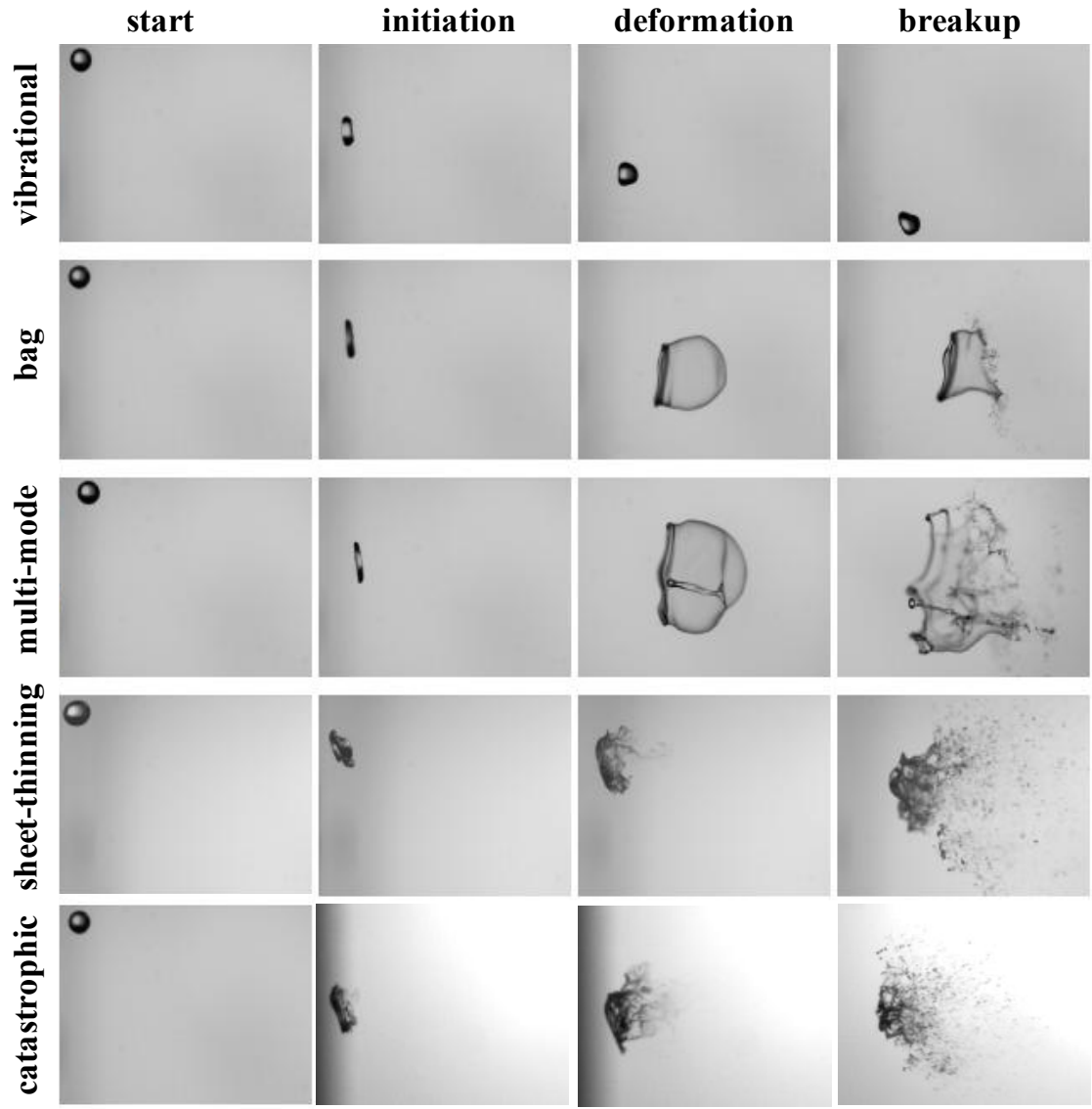

Figure 2.2: High speed images showing the five regimes of breakup for a low viscosity (inviscid) Newtonian drop in the presence of a horizontally flowing air stream, Guildenbecher et al. (2009).

The breakup process starts when the drop enters a disruptive flow field. The most commonly accepted breakup modes are shown in Fig. 2.1. Fig. 2.2 contains frames extracted from high speed videos. The following sections offer a qualitative description of these modes and some insights into the mechanisms involved.



### 2.3.1.1   Vibrational Breakup Mode

As *We* increases from zero, the first of the breakup modes is *vibrational*. See Fig. 2.3. Here the drop oscillates with a definite frequency and, if time permits breakup to occur, leads to the formation of a few fragments whose size is on the order of the original drop. Not many investigations have considered this regime because it does not guarantee breakup. In fact, Hsiang and Faeth (1992) argue that since the oscillations may either be stable or unstable breakup is not assured. Nevertheless, the importance of this regime lies in the fact that it precedes bag breakup where breakup is guaranteed.

The drag on the drop undergoing vibrational breakup is an important consideration. A simple force balance model, such as the one shown below, has been proposed, but it suffers from limitations.

$$F_D = \frac{1}{2} \rho_g U_{rel} C_D \frac{\pi d_{cro}^2}{4}. \tag{2.7}$$

Determination of $C_D$ is the major challenge as it requires a description of drop deformation versus time. Hsiang and Faeth (1992) found the maximum cross stream diameter to be

$$\left( \frac{d_{cro}}{d_0} \right)_{max} = 1 + 0.19 We^{0.5}. \tag{2.8}$$



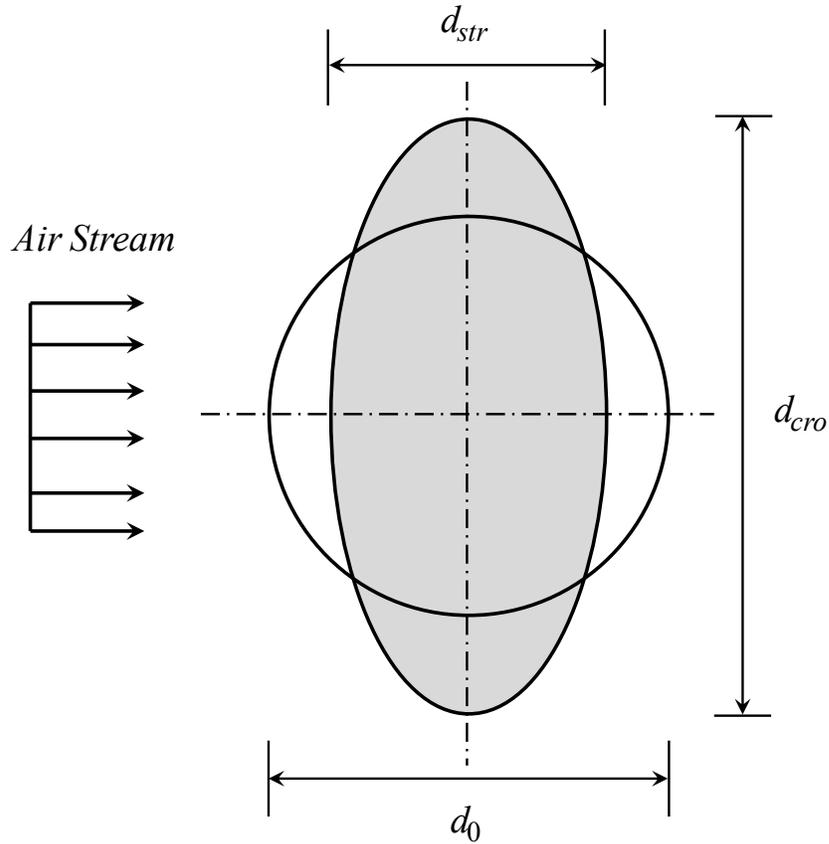

Figure 2.3: Vibrational mode of breakup (Guildenbecher, 2009). $d_0$ is the initial drop diameter, $d_{cro}$ is the cross stream length and $d_{str}$ is the streamwise length.

Pilch and Erdman (1987) and Gelfand (1996) used equation (2.6) using an average drag coefficient and $d_0$ where $\overline{C_D}$ was given by,

$$\overline{C_D} = 1.6 + 0.4\,Oh^{1.08}We^{0.01}$$
$$1000 < We < 162000,\ Oh < 0.44,\ 0.95 < Ma < 1.63. \tag{2.9}$$

This model fails to predict the instantaneous acceleration correctly. The Taylor Analogy Breakup (TAB) and drop deformation and breakup (DDB) models have since been developed and have been shown to predict drop behavior much more accurately.



### 2.3.1.2   Bag Breakup Mode

As *We* increases beyond the range for vibrational breakup, *bag breakup* will occur. At this stage the aerodynamic forces play a sufficiently strong role to guarantee breakup into fragments that are much smaller than the original drop. The first step in bag breakup is the presence of an unequal pressure distribution around the drop, with the result that the drop deforms into an oblate spheroid. Further deformation transforms the oblate spheroid into a thin hollow bag attached to a thick toroidal rim. The bag disintegrates, first owing to instabilities on its surface. Its disintegration results in a large number of very small fragments whose diameters are on the order of the bag thickness. Bag disintegration is followed by breakup of the toroidal rim. This breakup is also due to instabilities; it results in a smaller number of larger fragments and their diameters are on the order of the rim thickness. All these stages are illustrated in Fig. 2.4.

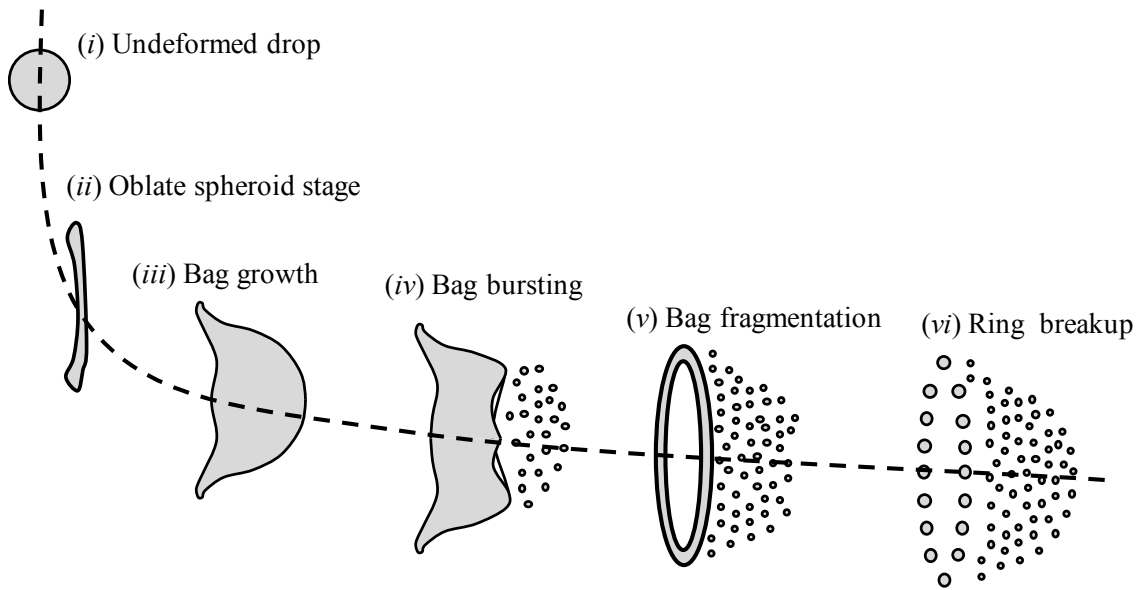

Figure 2.4: Stages of bag breakup (*i*) undeformed drop (*ii*) oblate spheroid stage (*iii*) bag growth (*iv*) bag bursting (*v*) bag fragmentation (*vi*) ring breakup.



The toroidal ring contains approximately 60% of the original volume (Chou and Faeth, 1998). Because the rim is much thicker than the bag fragments formed from the toroidal ring disintegration are much larger than those formed from fragmentation of the bag. In fact, Chou and Faeth (1998) reported the mean diameter of the ring breakup fragments to be 30% of the original drop diameter, while the mean diameter of bag breakup fragments were approximately 4% of the original drop diameter.

Prior to that, Hwang *et al*. (1996) noted that the bag first broke into ligaments that were aligned with the flow direction—this ruled out breakup for the bag. Liu and Reitz (1997) postulated that small holes formed in the bag leading to its eventual rupture. This we now know is due to thickness modulations caused by Rayleigh-Taylor instability. Villermaux and Bossa (2009) reported findings on falling raindrops, in which they put forth a model for bag growth. Chapter 4 of this document discusses detailed modeling aspects of this process using Villermaux and Bossa's model. The study of this breakup is the focus of this thesis.

### 2.3.1.3  Multimode Breakup Mode

This type of breakup has also been called bag-and-stamen breakup. Cao *et al.* (2001) termed this the "dual bag breakup" mode, but it also can be classified under this category.

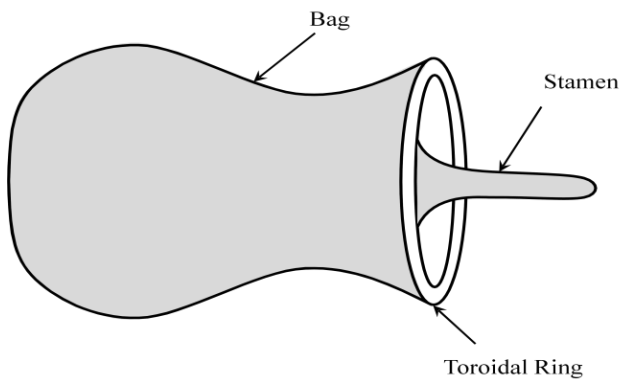

Figure 2.5: Bag and stamen breakup.



Dai and Faeth (2001) suggest that bag/plume breakup occurs for $\sim18 < We < \sim40$ and plume/sheet-thinning occurs for $\sim40 < We < \sim80$, both with $Oh < 0.1$. Regardless of title, it has three distinct features. In the order they occur they are: breakup of the bag, breakup of the rim, and breakup of the stamen (the stamen is a protrusion oriented in the direction opposite to the flow, Fig. 2.5). Because there are three breakup steps we obtain fragments of multiple sizes.

### 2.3.1.4   Sheet Thinning Breakup Mode

In sheet stripping breakup (now referred to as sheet-thinning), the outer periphery of the drop is continuously removed from its surface. As this periphery disintegrates it forms droplets. This continues until complete disintegration is achieved (see Fig. 2.6).

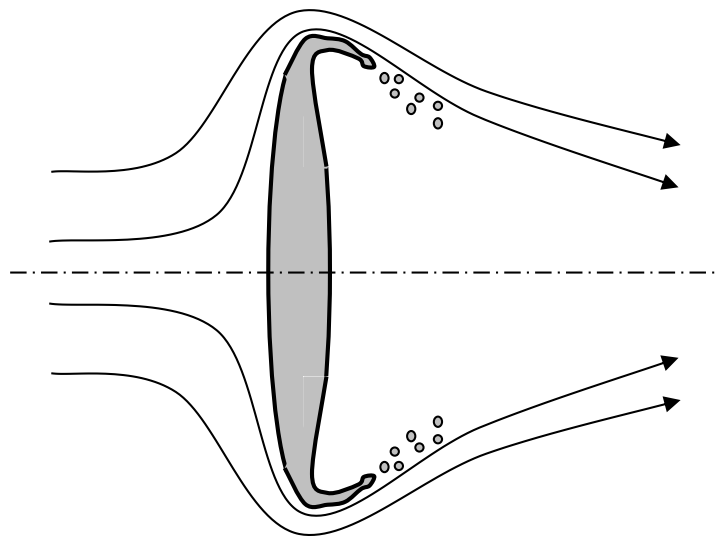

Figure 2.6: Sheet-thinning breakup (Guildenbecher, 2009).

One of the first explanations for sheet thinning breakup was by Ranger and Nicholls (1969) who posited that the mechanism was similar to an unstable boundary layer being stripped from the drop surface. This mechanism is also referred to as boundary layer stripping or shear stripping. Chou and Faeth (1997), with their experiments on viscous



drops having *Oh* < 0.1, supported this hypothesis. It was subsequently disproved by DNS calculations with the sheet thinning mechanism proposed by Reitz and co-workers found to be more appropriate. In sheet thinning the ambient phase inertia causes the periphery of the drop to be turned in the direction of the flow, which then forms a sheet. This sheet undergoes further deformation and eventually breaks into individual fragments.

### 2.3.1.5   Catastrophic Breakup Mode

As shown in Fig. 2.7, the drop almost instantaneously breaks apart. The drop surface is corrugated by waves of large amplitude and long wavelengths. They form a small number of large fragments that in turn break up into even smaller units. This process happens in a much shorter time frame than the other breakup modes. Wierzba and Takayama (1988) used holographic interferometry to study this type of breakup and confirmed the presence of drops being stripped from a large portion of the surface early in the breakup process, rather than just the drop periphery as hypothesized to occur in the sheet-thinning regime.

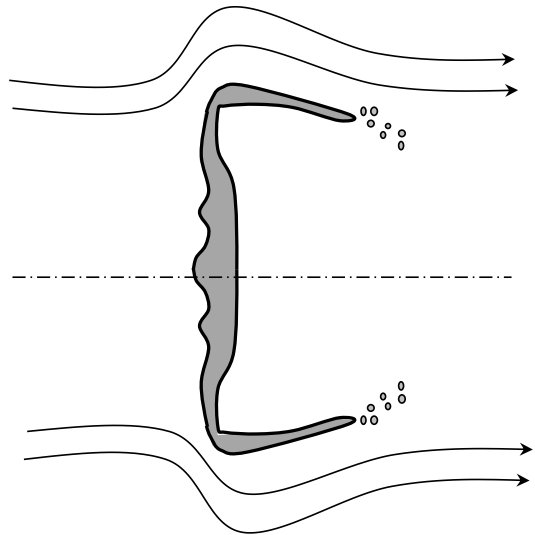

Figure 2.7: Catastrophic breakup.

At later times they observed the drop core breaking into large fragments, which in turn underwent stripping breakup. Patel (1981) has attributed this to a combined Raylor-



Taylor/Kelvin-Helmholtz instability, with the former taking place at the core where the liquid mass is accelerated perpendicular to the direction of the flow and the latter at the edge where liquid is interacting directly with the ambient. Theofanous (2011) has argued though that this regime is an artifact of shadowgraphy in earlier characterizations.

## 2.4    Regime Boundary

Different types of breakups are seen for varying flow conditions,. Establishing these boundaries is important from the point of view of studying atomization behavior. The transition, although not a continuous one, has been assumed as such in order to simplify. Various empirical correlations have been out to mark the critical $We_c$ for transition. Note that some of these are for shock tube experiments, like that of Brodkey (1967), but general applicability to other methods is assumed.

Brodkey (1967) proposed the following correlation, which was confirmed by Pilch and Erdman (1987) for $Oh < 10$,

$$We_c = We_{cOh \to 0} \left( 1 + 1.077\, Oh^{1.6} \right). \tag{2.10}$$

We use this correlation in Chapter 4 for our comparison. Most studies have found the transition between two modes to be a function of $We$ and $Oh$ and independent of other parameters, such as $\varepsilon$ or $Re$. For $Oh < 0.1$ the transition $We$ are as shown in Fig. 2.2. Since, the transition happens over a range of $We$ different authors, Pilch and Erdman (1987) and Hsiang and Faeth (1995) amongst others, have reported various values.

Here, $We_{cOh \to 0}$ is the critical $We$ at low $Oh$. Gelfand (1996) reviewed mostly Russian works and proposed:

$$We_c = We_{cOh \to 0} \left( 1 + 1.5\, Oh^{0.74} \right). \tag{2.11}$$



Cohen (1994) assumed that in the absence of drop viscosity, the kinetic energy imparted by the ambient flow to the drop is equal to the increase in surface energy. An extra energy term was added to account for the drop viscosity, therefore increasing the kinetic energy needed to cause breakup. He came up with the following correlation:

$$We_c = We_{cOh \to 0} \left( 1 + cOh \right). \tag{2.12}$$

These correlations are compared in Fig. 2.8. They do not agree at $Oh > 3$. Fig. 2.9 is important in the context of delineating regime boundaries. It is a regime map which shows the various modes along with their limiting values. We see that for $Oh<0.1$ the regime boundaries are flat and there is no effect of $Oh$. In contrast, as we go to higher values of $Oh$ these boundaries become more curved and the effect of $Oh$ sets in.

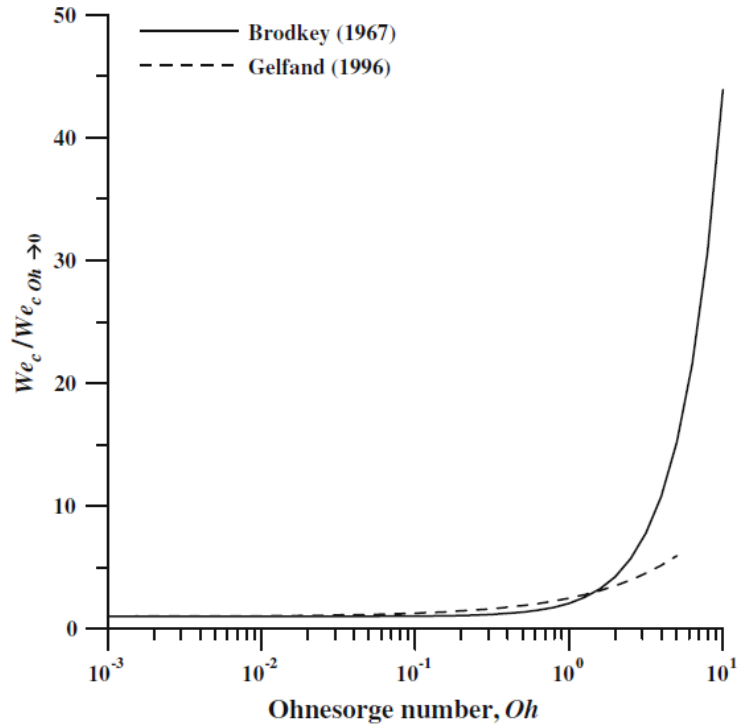

Figure 2.8: Comparison of correlations of Gelfand (1996) and Brodkey (1994).



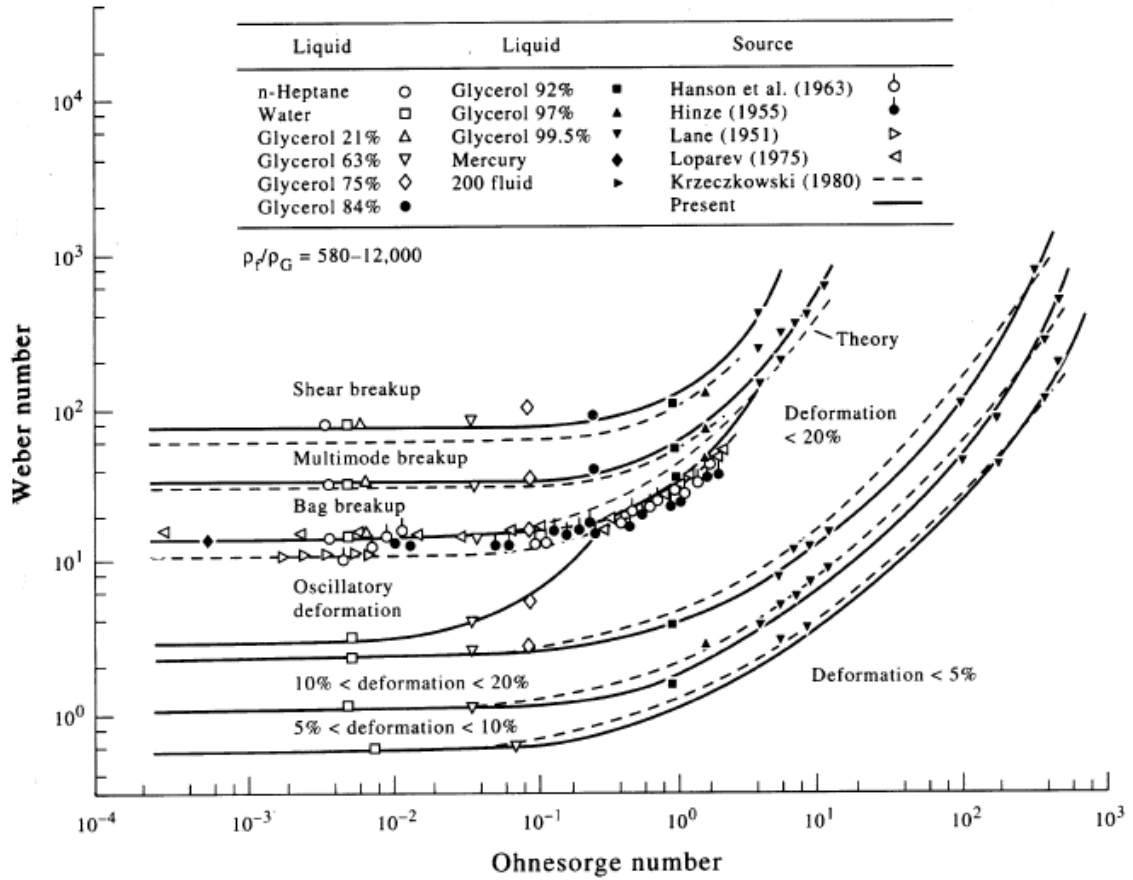

Figure 2.9: *We* at transition (Hsiang and Faeth, 1995). Drop deformation and breakup due to shock wave and steady disturbances.

## 2.5    Breakup Times

The time required for various deformations to take place is an important marker of the topological change within the drop. The two most commonly referred to times are the initiation time and the total breakup time. These are described in detail in the sections that follow.



### 2.5.1   Initial Breakup Time

The initial breakup time is defined as the interval required for a drop to deform beyond the oblate spheroid shape after it has been exposed to aerodynamic loading. This time is important because it indicates the point when the models of a deforming ellipsoid, discussed in the section on deformation and vibration, are no longer valid. The correlations proposed by Pilch and Erdman (1987), as given by Gelfand (1996), and Hsiang and Faeth (1992), are given below.

$$T_{ini} = 1.9 \left( We - We_c \right)^{-0.25} \left( 1 + 2.2 Oh^{1.6} \right)$$
$$We < 10^4, Oh < 1.5. \tag{2.13}$$

$$T_{ini} = \frac{1.6}{1 - Oh / 7}$$
$$We < 10^3, Oh < 3.5. \tag{2.14}$$

$$T_{ini} = 1.4 \left( 1 + 1.5 Oh^{0.24} \right)$$
$$We \approx We_c, Oh < 4.0. \tag{2.15}$$

### 2.6   Analytical Modeling

One of the easiest modes of breakup to study mathematically is the vibrational mode. As mentioned earlier, when compared to simple empirical models as proposed by Pilch and Erdman (1987), the TAB model proposed by O'Rourke and Amsden (1987) has been found to work well. This model uses an analogy between an oscillating drop and spring-mass-damper system. The spring force, external driving force, and dampening are respectively analogous to surface tension, aerodynamic force, and drop viscosity. Breakup is assumed to occur when $d_{str}$ tends to zero.



The TAB model has been used to describe secondary atomization by O'Rourke and Amsden (1987), Liu and Reitz (1993), Hwang *et al.* (1996), Tanner (1997), Lee and Reitz (1999), and Park *et al.* (2002). Over a period of time some shortcomings have come to the fore. For example, breakup was assumed to occur instantaneously, while it is now known that it is a gradual process. In addition, drop trajectories were found to be inadequately described. Finally, the breakup criterion was somewhat arbitrary and experimental data, such as equation (2.10), indicate that the critical deformation was actually a function of *We*.

There are three known variants of the TAB model.

The NLTAB-model is a nonlinear variant of the TAB-model developed by Schmehl (2002) and is based on the experimental observation that the droplet deformation in air flows can be represented by spheroidal shapes (oblate and prolate). Thus, it accounts for large deformations which were previously not taken into consideration in the TAB model. The NLTAB-model describes the droplet deformation dynamics in terms of the non-dimensional equator coordinate $Y$, which is defined as the ratio of the cross-stream diameter to the spherical diameter, $Y = {d_{cro}}/{d_0}$.

The Enhanced Taylor Analogy Breakup (ETAB) was developed by Tanner (1997) and simulates the disintegration process as a cascade of drop breakups. The breakup criterion is determined by Taylor's linear drop deformation dynamics and the associated drop breakup condition. Breakup occurs when the normalized drop distortion exceeds a critical value. According to this model the drop deformation is described by the forced damped harmonic oscillator.

The DDB model was originally proposed by Ibrahim *et al.* (1993) and is applicable to high *We* drops. It is based on the motion of center of mass of the half droplet and assumes that the droplet is distorted by pure extensional flow. It uses conservation of energy for a



distorting droplet with drop deformation is calculated by equating the rate of change in kinetic and potential energies to the work done on the drop due to pressure and viscous forces. Breakup is assumed to occur when both kinetic and viscous forces are negligible.

Hwang *et al*. (1996), Liu and Reitz (1997), Park *et al*. (2002), and Pham and Heister (2002) have used this model. It does have some shortcomings, such as that noted by Park *et al*. (2002) including predicting instantaneous breakup without deformation when *We* is less than 19. This is clearly unrealistic. Nevertheless, currently both models are used in industrial spray simulations, each with their own advantages and limitations.

A number of authors have proposed improvements to the TAB and DDB models. The breakup model of Lee and Reitz (1999), in which a Kelvin-Helmholtz instability model is employed (with limited success) is one such case. Some hybrid models are also used to simulate the simultaneous effect of breakup due to unstable wave growth and breakup due to aerodynamic deformation, such as is expected in the catastrophic breakup mechanism.

In the bag breakup regime relatively few models have been developed owing to the complexity of flow. Rayleigh-Taylor based mechanisms have been put forth, but they failed to capture the process in detail. As an example, Tarnogrodzki (1993) developed a model based on the duct flow solution. Drag coefficients for a disk and sphere were used to approximate the dynamic pressure acting on the drop and breakup was assumed to occur when the radial motion of the drop ceased. The model was able to calculate $We_c$ to within the correct order of magnitude, but predicted that $We_c$ continuously decreased with $Oh$, even for $Oh < 0.1$ which was in contrast to what is observed in experimental findings.

## 2.7    Fragment Size Distributions

Fragment size distribution is one of the most important features of any atomization process. Combustion, agriculture, and all applications mentioned earlier require



knowledge of the fragment size for optimal design of an atomizer. So far, fragment sizes have been only rarely measured.

Fragment size distributions are often reported as fractions of a mean, or characteristic, diameter. The mean, or characteristic, drop diameter can be described in many ways. A representative diameter ($D_{pq}$) is,

$$D_{pq} = \left[ \frac{\int_0^\infty D^p f_0(D) dD}{\int_0^\infty D^q f_0(D) dD} \right]^{\frac{1}{p-q}}.$$

(2.16)

where $p$ and $q$ are positive integers and $f_0(D)$ is the number pdf. The arithmetic mean diameter ($D_{10}$), the volume mean diameter ($D_{30}$), and the Sauter mean diameter ($D_{32}$) are commonly used.

Hsiang and Faeth (1995) results for $Oh < 0.1$ have shown that the bag and the bag and stamen regime fragment mass median diameter (MMD) and $D_{32}$ are related by $MMD/D_{32}$. Empirical correlations, such those given below have also been suggested by Hsiang and Faeth (1995).

$$We_{32} = c_1 \varepsilon^{0.25} Oh^{0.5} We^{0.75}$$
$$We < 1000, \, Oh < 0.1, 580 < \varepsilon < 1000$$

(2.17)

$$We_{32} = \frac{\rho_g D_{32} U_{rel}^2}{\sigma} \quad \text{and} \quad c_1 = 6.2.$$

The Maximum Entropy Formalism (MEF) has been considered for describing the drop breakup diameter distribution. Babinsky and Sojka (2002) provide a thorough discussion of the development and application of the MEF to sprays applications. Later, Li *et al.* (2005) noted that the MEF is applicable to isolated systems in thermodynamic equilibrium and many sprays do not meet these requirements. They therefore proposed a change which matched predictions with experimental data, but necessitated the use of



more characteristic diameters which is not always possible apriori. Cousin *et al*. (1996) proposed the use of linear stability theory to predict one characteristic fragment diameter. However, no theoretical method exists to predict the second characteristic diameter, so either experimental results or ad hoc assumptions are required. Dumouchel (2006) also tried to solve the problem of too few known characteristic diameters, but ended up adding more complexity. Therefore, the validity of MEF is still open to debate.

2.8    Computational Work

The two main computational techniques have been direct numerical simulation ( DNS) and volume of fluid (VOF) methods. Each has succeeded to varying extents, but extracting all details of the breakup process is very difficult. However, DNS, as used by Thefanous (2011, 2012) has offered significant advancements in the current understanding of the break up phenomena. Very recently, DNS and VOF were used by Jalal and Mehravaran (2012). They studied disintegration of falling liquid droplets in quiescent media for different Eötvös numbers. Three simulations with different Eötvös numbers were performed. The VOF method based on octree meshing was used which provided a considerable reduction in computational cost. The exact topology of the deforming drops did not seem to have been captured. Nevertheless, this work is the first one to perform 3-D simulations as opposed to using a 2–D or axisymmetric domain.

Igra and Takayama (2001) and Igra *et al*. (2002) used a VOF method to track the interface and included surface tension effects. Their studies revealed formation of the bag in the direction opposite to that observed in experiments. They concluded that this could have been due to incorrect initial conditions.

One of the important computational studies of note is that of Han and Tryggvason (1999, 2001) which aimed to address many of the shortcomings of earlier studies. A front tracking/finite difference method was used to solve the axi-symmetric Navier-Stokes



equations. The axi-symmetric assumption allowed for the simulation of a spherical drop rather than a 2D cylinder. Both bag and sheet-thinning type structures were observed, but transitional *We* values did not match those seen in experiments. This may be due to the fact that calculations were performed for ε <10 to reduce computational cost (most experiments are performed for liquid drops in ambient gas environments where the density ratio is much higher).

The Lattice-Boltzmann approach, as presented by Sehgal *et al.* (1999), has also been used. It failed to estimate transition *We* accurately so more work needs to be done. Other methods such as moving-particle semi-implicit (MPS) have also been used. They also were unable to capture all the drop dynamics accurately. Fakhri and Rahimian (2009) also used the Lattice-Boltzman approach, but their study too was 2-D/axisymmetric in nature. Obviously, more work in light of experimental findings needs to be done to confirm validity of numerical schemes.

2.9    Summary

The present literature review summarizes the work done so far in studying secondary atomization of Newtonian and non-Newtonian liquids. Five breakup regimes were identified: vibrational, bag, bag and stamen/mutlimode, sheet thinning, and catastrophic. The regime boundaries and their underlying mechanisms were then discussed. The salient features of each of these breakup modes from the standpoint of fragment distribution were also highlighted. The importance of breakup times, i.e., initiation time and total breakup time, was then described. Some empirical correlations useful for engineers were also quoted. After a discussion of fragment size distributions, a review of analytical and computational works was given.

From the review it was concluded that a lot of experimental work has been done in the case of Newtonian drops, and there is a fair amount of consistency in the findings of different researchers. However, only a few studies have considered non-Newtonian drop



breakup. The breakup mechanisms observed were similar to those reported for Newtonian drops. However, there is still not enough data to determine common characteristics or processes for non-Newtonian drop secondary breakup. The significant weaknesses in current Newtonian secondary breakup studies are as follows:

1. There is no well-established theoretical model for bag breakup of viscous Newtonian drops.

2. The underlying physical mechanisms governing bag breakup need to be verified.

3. Detailed study of fragment size distributions in the bag breakup regime for Newtonian fluids has not been reported.

4. Theoretical determination of the regime boundary for viscous drops.

The present thesis addresses these four inadequacies.



## CHAPTER 3. EXPERIMENTAL SETUP

This chapter summarizes the secondary atomization test facility and properties of the liquids used in the characterization of drop breakup. Also included are comments on measurement methods and corresponding experimental uncertainties. Much of this text is taken from Lopéz (2010) and Synder (2011) since the same test facility was used. The details of the PDA system to make drop size measurements are also given. The experimental facility is shown as Fig. 3.1. It consists of an air supply sub-system, a liquid sub-supply (drop generator and traverse), a high speed imaging system and a Phase Doppler Anemometry (PDA). Each is described below.

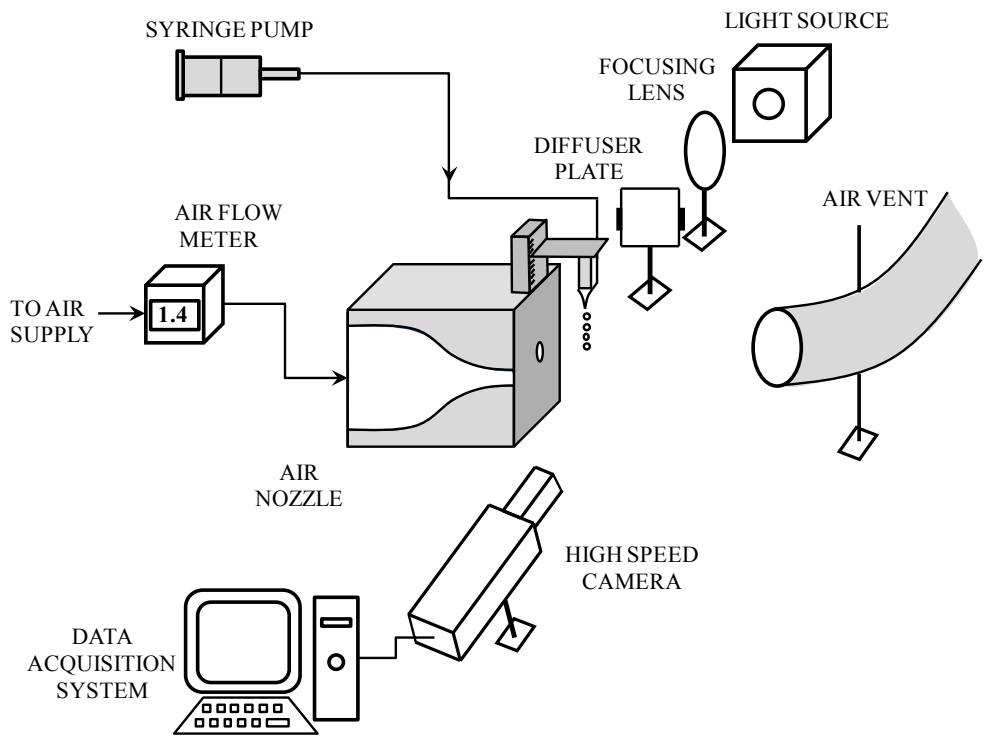

Figure 3.1: Experimental setup.



### 3.1  Apparatus

### 3.1.1  Air Supply

Air was used to create the aerodynamic disruptive flow field for all experimental runs. Compressed air at 600 KPa and room temperature was supplied to the facility. First, the air passed through a shutoff valve and a Micro Motion F-Series Coriolis flow meter. A needle valve was used for flow rate control. The air then flowed through a bypass unit that allowed insertion of a TSI, Inc. oil drop generator (model 9307-6) that was used for LDA and PIV seeding, before being routed to the air nozzle through a flexible 1.27 cm diameter stainless steel tube.

### 3.1.2  Liquid Supply

The liquid to be atomized was contained in an Alloy Products Corp. 316 L stainless steel pressure tank (S/N S501647-1-013). The liquid was routed to a syringe pump using 6 mm diameter PVC tubing. The syringe pump outlet was connected to a plastic block that housed a square cut syringe needle. The needle allowed individual drops to fall into the air stream as a result of gravity.

The air nozzle was made from clear acrylic and was designed to produce a nearly uniform velocity profile at its exit. It has a 15 cm diameter inlet section and houses a Plascore Inc. polycarbonate honeycomb (4 mm cell diameter and 2.54 cm cell length) to suppress large scale eddies. Downstream of the honeycomb is a fine wire mesh (0.05 mm wire diameter and 0.07 mm spacing between wires), which produced homogeneous turbulence that quickly dissipated. The nozzle ends with a converging outlet section of 2.54 cm diameter, as shown in Fig. 3.1. This results in a steady, nearly one dimensional, laminar flow field for drop breakup. The air velocity profile measured using PIV and LDA agreed to within experimental uncertainty. These details can be found in Guildenbecher (2009).



### 3.1.2.1 Drop Generator

The key part in the drop generator was an EFD Inc. stainless steel syringe tip with an inner diameter of 0.25 mm. Liquid entered the nozzle through the side and exited through the syringe tip.

### 3.1.2.2 Traverse

The air nozzle and drop generator were mounted on a Velmex three-dimensional traverse to facilitate accurate positioning of drops falling into the air jet. The traverse was composed of three UniSlide electro-mechanical motor driven assemblies and a NF90 series programmable stepping motor controller. The UniSlide assemblies had 2024-T3 type hardened aluminum sliders with bonded bearing pads of a PTFE composite formulation. They converted rotary to linear motion through a precision roll-formed lead screw having a spatial resolution of 5 μm per step.

### 3.1.3   High Speed Imaging System

Time-dependent drop morphology and breakup times were determined using a Vision Research Phantom 7.1 high speed digital camera and a Nikon lens of 105 mm focal length. Images of $800 \times 600$ pixels were recorded at 4700 fps using the Phantom Camera Control Software (version 9.1.663.0-C PhCon: 663). A Kratos 1000 W Xe arc lamp (model LH151N) was used as the illumination source. It produced a collimated beam, which was reflected by a dichroic mirror that filtered out the IR and UV components. A plano-convex lens was used to focus the beam. A calibration image with a 5 mm square grid was also part of the imaging system. It was used to determine axial and radial locations with respect to the air nozzle exit, and also enabled determination of drop sizes directly from the images by providing a mapping relationship between camera pixels and distance. The optical arrangement is as shown in Fig. 3.1.



### 3.1.4    PDA (Phase Doppler Anemometry)

Déchelette (2010) gives a description of the apparatus that which can be referred to for specific details. The salient features of the setup are highlighted here. The Dual PDA manufactured by Dantec Dynamics is primarily used to make the drop size measurements. A transmitter with a 310 mm focal length lens and a receiver equipped with a 310 mm focal length lens, along with receiver mask C and a 200 μm wide slit was used. All photomultiplier (PMT) voltages were set to 1200 V to ensure burst signals on all four channels. The Dual PDA was used in its 2D configuration. The spherical validation was between 90 and 95% in all cases. The scattering angle was set to $30^0$, the signal-to-noise ratio (SNR) to 0 dB, and the signal gain to 20 dB to ensure that drop sizes are captured. The use of a $f$ = 310 mm transmitter lens and a $f$ = 310 mm receiver lens leads to a measurement volume having $\delta x$ = 0.5 mm, $\delta y$ = 0.5 mm and $\delta z$ = 30 mm. For the Non-Newtonian drops the fragments consisted primarily of stretched ligaments. Since, this would lead to incorrect measurements the Non-Newtonian drop sizes were not recorded. A Fiber PDA with a transmitter of 1000 mm focal length lens and a receiver equipped with a 310 mm lens, along with receiver mask B was also used to confirm some of the findings made by the Dual PDA. This helped in increasing the measurement volume and increased data sample rates.

Sample sizes of the drops varied from 5000 to 10000 depending on the viscosity of the solution. These were enough to ensure statistical significance. As the viscosity of the solution increased the data rates dropped as fewer fragments were being produced. The measurements were made at a location 30 cm downstream of the air nozzle exit. Since the bag fragments before the rim, this distance corresponded to the point where the rim had just about undergone breakup. This time duration is short enough to ensure that the bag fragments do not undergo sufficient evaporation.



## 3.2    Solutions Tested

Two types of liquids were tested inviscid and viscous Newtonian liquids. The tables below give details of the Newtonian liquids tested.

Table 3.1: Liquid properties for Newtonian fluids tested.

| Solution (*weight %*) | Density $\rho_l\,(kg/m^3)$ | Surface Tension $\sigma\,(N/m)$ | Dynamic Viscosity $\mu_l \times 10^4\,(Pa-S)$ |
|---|---|---|---|
| D.I. Water | $997 \pm 2$ | $0.0710 \pm 0.008$ | $8.94 \pm 1$ |
| Ethanol | $800 \pm 4$ | $0.0240 \pm 0.005$ | $16 \pm 2$ |
| 40% Glycerine | $1100 \pm 8$ | $0.0662 \pm 0.003$ | $30 \pm 2$ |
| 50% Glycerine | $1130 \pm 7$ | $0.0651 \pm 0.007$ | $72 \pm 4$ |
| 63% Glycerine | $1162 \pm 10$ | $0.0648 \pm 0.003$ | $108 \pm 6$ |
| 70% Glycerine | $1185 \pm 4$ | $0.0640 \pm 0.010$ | $356 \pm 9$ |

## 3.3    Parameters Measured and Non−Dimensional Numbers Used

A dimensional analysis considering the dominant forces leads to the following non-dimensional numbers for Newtonian liquids,

$$We = \frac{\rho_g U_{rel}^2 d_0}{\sigma}. \tag{3.1}$$

$$Re = \frac{\rho_l U_{rel} d_0}{\mu_l}. \tag{3.2}$$

$$Oh = \frac{\mu_l}{\sqrt{\rho_l d_0\,\sigma}}. \tag{3.3}$$



where $\rho_g$ is the density of the gas, $\rho_l$ the density of the liquid, $\sigma$ the surface tension, $d_0$ the initial drop diameter, $\mu_g$ the viscosity of the gas, $U_{rel}$ the mean air velocity. Droplet geometry measurements were made using the open source software IMAGE J (NIH) and Cine Viewer 663. Various parameters connected to droplet deformation are measured. Fig. 3.2 shows the various parameters quantified: the thickness of the rim at any instant, $h(t)$, maximum bag extent $\alpha(t)$, and the radial extent $2R(t)$. Three runs for each *We* corresponding to each *Oh* were taken. The values represent the arithmetic mean of measurements made from these videos.

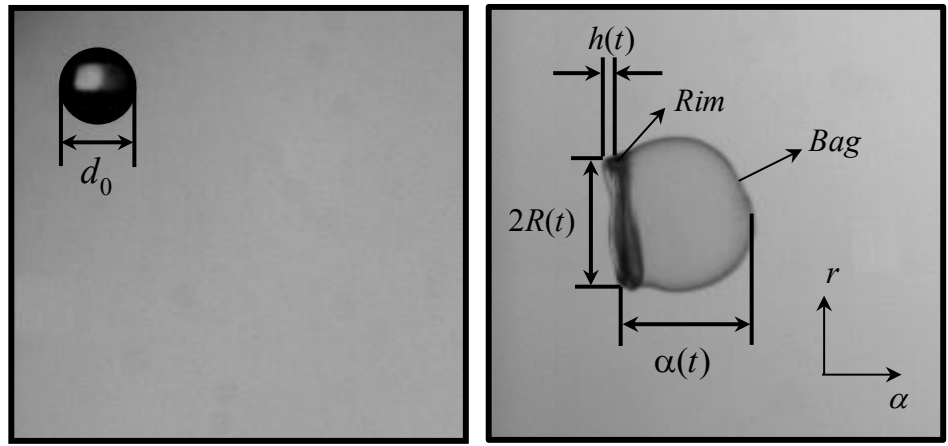

Figure 3.2: Various quantities associated with bag geometry measured in this study.



Table 3.2 gives the test conditions corresponding to bag breakup of Newtonian drops.

Table 3.2: Test conditions.

| Test Conditions (at 293 K , 1 *atm* ) | | | | | |
|---|---|---|---|---|---|
| Solution (% *weight*) | Drop Diameter $d_0\,(mm)$ | We | Oh | Re | Density Ratio $\rho_l/\rho_g$ |
| D.I. Water | $2.6 \pm 0.10$ | 14 to 16 | 0.002 | – | 828 |
| Ethanol | $1.9 \pm 0.1$ | 12 to 14 | 0.007 | – | 655 |
| 40% Glycerine | $2.1 \pm 0.1$ | 13 to 14 | 0.010 | – | 912 |
| 50% Glycerine | $2.3 \pm 0.03$ | 12 to 15 | 0.015 | – | 934 |
| 63% Glycerine | $2.0 \pm 0.05$ | 12 to 15 | 0.034 | – | 964 |
| 70% Glycerine | $2.2 \pm 0.05$ | 12 to 14 | 0.058 | – | 980 |
| Air | – | – | – | 1320–2820 | 1 |

## 3.4 Experimental Uncertainties

Uncertainty in the experimental results arises due to factors such as image processing, initial air flow conditions, and liquid characterization. The uncertainty in these quantities contributes to uncertainties in the dimensionless parameters *We*, *Oh*, and *T*. In this section, the uncertainties for these groups, and for other quantities, are discussed.

The uncertainty in the air flow rate is due to the flow meter reading. The flow meter reads to the nearest $\pm 0.01$ kg/min. This gave a relative uncertainty of less than $\pm10\%$.

### 3.4.1 Uncertainty in Relative Velocity ($U_{rel}$)

Since the relative velocity plays an important role in the aerodynamic disintegration process, determining the uncertainty in $U_{rel}$ is vital in evaluating results. Both LDA and



PIV techniques (Guildebecher, 2009) were employed to determine the air flow velocity and agreement in their centerline measurements was found to be within ±6%. The uncertainty in air velocity was measured in terms of the turbulent intensity and was observed to be ±3%. $U_{rel}$ is taken to be equal to the measured air velocity, since the drops fall vertically so its uncertainty, $u_{U_{rel}}$ is also ±3%.

### 3.4.2   Uncertainty in Initial Drop Diameter ($d_0$)

Uncertainty in the initial drop diameter is due to uncertainty in locating the drop edge. This can be accomplished to within ±1 pixel. After using the appropriate pixel-to mm scale, the uncertainty in the drop diameter was estimated to be ±0.1 mm. Since most of the drops had an initial diameter of 4 mm, the relative uncertainty in drop diameter is approximately ± 2.5%.

### 3.4.3   Uncertainty in $We$

$$u_{We} = \pm \left[ \left( \frac{d_0}{We} \frac{\partial We}{\partial d_0} u_{d_0} \right)^2 + \left( \frac{U_{rel}}{We} \frac{\partial We}{\partial U_{rel}} u_{U_{rel}} \right)^2 + \left( \frac{\rho_g}{We} \frac{\partial We}{\partial \rho_g} u_{\rho_g} \right)^2 + \left( \frac{\sigma}{We} \frac{\partial We}{\partial \sigma} u_{\sigma} \right)^2 \right]^{\frac{1}{2}}. \quad (3.4)$$

In (3.4), quantities, $\dfrac{\partial We}{\partial d_0}$, $\dfrac{\partial We}{\partial U_{rel}}$, $\dfrac{\partial We}{\partial \rho_g}$, $\dfrac{\partial We}{\partial \sigma}$, are,

$$\frac{\partial We}{\partial d_0} = \frac{\rho_g U_{rel}^2}{\sigma}. \quad (3.5)$$

$$\frac{\partial We}{\partial U_{rel}} = 2 \frac{\rho_g U_{rel} d_0}{\sigma}. \quad (3.6)$$

$$\frac{\partial We}{\partial \rho_g} = \frac{U_{rel}^2 d_0}{\sigma}. \quad (3.7)$$



$$\frac{\partial We}{\partial \sigma} = -\frac{\rho_g U_{rel}^2 d_0}{\sigma^2}. \tag{3.8}$$

Typical results reveal $u_{We} \pm 11\%$ with $u_{We} \pm 15$. The uncertainty in the diameter is the largest source of uncertainty in $We$.

### 3.4.4   Uncertainty in $Oh$

$$u_{Oh} = \pm\left[\left(\frac{d_0}{Oh}\frac{\partial Oh}{\partial d_0}u_{d_0}\right)^2 + \left(\frac{\rho_l}{Oh}\frac{\partial Oh}{\partial \rho_l}u_{\rho_l}\right)^2 + \left(\frac{\sigma}{Oh}\frac{\partial Oh}{\partial \sigma}u_{\sigma}\right)^2 + \left(\frac{\mu_l}{Oh}\frac{\partial Oh}{\partial \mu_l}u_{\mu_l}\right)^2\right]^{1/2}. \tag{3.9}$$

Quantities $\dfrac{\partial Oh}{\partial d_0}, \dfrac{\partial Oh}{\partial \rho_l}, \dfrac{\partial Oh}{\partial \sigma}, \dfrac{\partial Oh}{\partial \mu_l}$ are given by,

$$\frac{\partial Oh}{\partial d_0} = -\frac{1}{2}\frac{\mu_l}{d_0\sqrt{\rho_l \sigma d_0}}. \tag{3.10}$$

$$\frac{\partial Oh}{\partial \rho_l} = -\frac{1}{2}\frac{\mu_l}{\rho_l\sqrt{\rho_l \sigma d_0}}. \tag{3.11}$$

$$\frac{\partial Oh}{\partial \sigma} = -\frac{1}{2}\frac{\mu_l}{\sigma\sqrt{\rho_l \sigma d_0}}. \tag{3.12}$$

$$\frac{\partial Oh}{\partial \mu_l} = \frac{1}{\sqrt{\rho_l \sigma d_0}}. \tag{3.13}$$

Using experimental values for the solution tested the maximum uncertainty in $Oh$ was found to be 14.3%.



### 3.4.5   Uncertainty in $T$

$$u_T = \pm \left[ \left( \frac{d_0}{T} \frac{\partial T}{\partial d_0} u_{d_0} \right)^2 + \left( \frac{U_{rel}}{T} \frac{\partial T}{\partial U_{rel}} u_{U_{rel}} \right)^2 + \left( \frac{t}{T} \frac{\partial T}{\partial t} u_t \right)^2 + \left( \frac{\rho_g}{T} \frac{\partial T}{\partial \rho_g} u_{\rho_g} \right)^2 \right.$$
$$\left. + \left( \frac{\rho_l}{T} \frac{\partial T}{\partial \rho_l} u_{\rho_l} \right)^2 \right]^{\frac{1}{2}} . \qquad (3.14)$$

Quantities, $\dfrac{\partial T}{\partial d_0}$, $\dfrac{\partial T}{\partial U_{rel}}$ $\dfrac{\partial T}{\partial t}$, $\dfrac{\partial T}{\partial \rho_g}$, $\dfrac{\partial T}{\partial \rho_l}$ are given by,

$$\frac{\partial T}{\partial d_0} = -\frac{t U_{rel}}{d_0^2} \sqrt{\frac{\rho_g}{\rho_l}} . \qquad (3.15)$$

$$\frac{\partial T}{\partial U_{rel}} = \frac{t}{d_0} \sqrt{\frac{\rho_g}{\rho_l}} . \qquad (3.16)$$

$$\frac{\partial T}{\partial t} = \frac{U_{rel}}{d_0} \sqrt{\frac{\rho_g}{\rho_l}} . \qquad (3.17)$$

$$\frac{\partial T}{\partial \rho_g} = \frac{t U_{rel}}{2 d_0 \sqrt{\rho_l \rho_g}} . \qquad (3.18)$$

$$\frac{\partial T}{\partial \rho_l} = -\frac{t U_{rel}}{2 d_0 \rho_l^{3/2}} \sqrt{\rho_g} . \qquad (3.19)$$

The maximum $u_T$ was found to be $\pm 7.25\%$.



# CHAPTER 4.  ANALYTICAL MODELLING

Theoretical treatment of bag breakup using the governing equations of mass and momentum conservation has not been reported in the literature.  Studies instead have been primarily empirical, or semi-empirical, in nature.  Restrictive simplifying assumptions were made to ease analytical treatment with the result that model predictions did not necessarily capture the physics underlying secondary breakup nor did they always match experimental results. A key contribution of this thesis is development of an analytical model for drop deformation and breakup that requires minimum empirical parameters.

As noted in Chapter 1 one of the main goals of this thesis is to develop a first principles analytical model to describe secondary breakup in the bag regime.  The model is to successfully predict behavior for inviscid and Newtonian liquids.

The inviscid model developed here follows that of Villermaux and Bossa (2009), changes will then be made for viscous effects.

## 4.1     Air Flow Field

We first model behaviour in stages 1 and 2, *i.e.* between the time the drop enters as a sphere until it ceases to be an oblate spheroid. Let, $U_r$ and $U_y$ be the velocities of air in the radial and streamwise directions.  Assuming the local air flow has the structure of a stagnation point, $U_y = a \dfrac{U_{rel} y}{d_0}$.  The value of $a$ is an indicator of the rate of stretching and $U_{rel}$ the mean velocity of the air jet as it exits the nozzle.



In the inviscid, incompressible, quasi-steady assumption conservation of air momentum and mass give,

$$\rho_g U_y \frac{\partial U_y}{\partial y} = -\frac{\partial p_g}{\partial y}. \tag{4.1}$$

$$\rho_g U_r \frac{\partial U_r}{\partial r} = -\frac{\partial p_g}{\partial r}. \tag{4.2}$$

$$r \frac{\partial U_y}{\partial y} + \frac{\partial (rU_r)}{\partial r} = 0. \tag{4.3}$$

From (4.3) we get $U_r = a \dfrac{U_{rel} r}{2d_0}$. Using (4.1) we may write,

$$p_g(r) = p_g(0) - \frac{a^2 U_{rel}^2}{8d_0^2} r^2. \tag{4.4}$$

where, $p_g(0) = \dfrac{\rho_g U_{rel}^2}{2}$ the stagnation pressure at $r = y = 0$.

## 4.2    Liquid Flow Field

To resolve the liquid flow field we write the Navier-Stokes equations in cylindrical coordinates,

$$\rho_l \left( \frac{\partial u_l}{\partial t} + u_l \frac{\partial u_l}{\partial r} \right) = -\frac{\partial p_l}{\partial r} + \mu \left[ \frac{1}{r} \frac{\partial}{\partial r} \left( r \frac{\partial u_r}{\partial r} \right) - \frac{u_r}{r^2} \right]. \tag{4.5}$$

$$r \frac{\partial h}{\partial t} + \frac{\partial (r u_r h)}{\partial r} = 0. \tag{4.6}$$



In (4.5) we note that $\mu = 0$ corresponds to the case of an inviscid drop. $u_r(r,t)$ is the velocity field inside the drop and global mass conservation gives,

$$h(t) = \frac{d_0^3}{6R^2(t)}.$$  (4.7)

This yields a velocity field of,

$$u_r(r,t) = r\frac{\dot{R}}{R}.$$  (4.8)

Substituting (4.8) in (4.5) we see that even for viscous drops the viscous stress terms cancel out. The viscous contributions enter the problem through the boundary conditions. First we consider the normal stress balance across the fluid interface, which is written as,

$$\sigma\kappa = T_{rr}(l) - T_{rr}(g).$$  (4.9)

Here, $T_{rr}(l)$ and $T_{rr}(g)$ represent the normal stress components associated with the liquid and the surrounding air and are given by $-p_l(r) + 2\mu_l\left(\frac{\partial u_r}{\partial r}\right)$ and $p_g(r)$, respectively, at a given radial location $r$. Specifying (4.9) at $r = R(t)$ yields,

$$p_l(R) = p_g(R) + \sigma\kappa + 2\mu_l\left(\frac{\partial u_r}{\partial r}\right).$$  (4.10)

where $\sigma$ is the surface tension at the interface, $\kappa$ is the curvature of the interface at $r = R(t)$, $\mu_l$ is the liquid dynamic viscosity. Using (4.4) in (4.10) we get,

$$p_l(R) = p_g(0) - \rho_g\frac{a^2U^2}{8d_0^2}R^2 + \frac{2\sigma}{h} - \frac{2\mu_l}{R}\dot{R}.$$  (4.11)

In (4.11), $\kappa$ evaluates to $\left(\frac{h(t)}{2}\right)^{-1}$ owing to the rounded corners of the liquid disk. The second boundary condition, *viz*, the tangential stress balance is inconsequential as the stresses in that direction on the interface are negligible.



## 4.3    Transition *We* and Radial Deformation, $\phi(t)$

Integrating the momentum equation (4.5) between $r = 0$ and $r = R(t)$ then using (4.8) and

non dimensionalising the result using $\phi = \dfrac{R}{d_0/2}$, $T = \dfrac{t}{\tau}$, we obtain,

$$\ddot{\phi} - 4\left(\frac{a^2}{16} - \frac{6}{We}\right)\phi + \frac{8Oh}{\sqrt{We}}\frac{\dot{\phi}}{\phi^2} = 0. \tag{4.12}$$

*We*, *Oh* are as defined in Chapter 2 overdots represents differentiation with respect to *T*, a

notation which shall be consistently used in the text hereafter. $\tau$ is the time scale typical

of the process of secondary atomization, and is given by,

$$\tau = \frac{d_0}{U_{rel}}\sqrt{\frac{\rho_l}{\rho_g}}. \tag{4.13}$$

$\tau^{-1}$ is particularly significant and denotes the frequency of oscillation when $Oh = 0$.

(4.12) is a second order nonlinear equation known as Lienard's equation and is similar to

a harmonic oscillator with damping. The coefficient of $\dot{\phi}$, $\left(\dfrac{8Oh}{\sqrt{We}}\right)\dfrac{1}{\phi^2}$, constitutes the

non-linear term which adds to damping even for small *Oh*. Equation (4.12) as such

cannot be solved analytically, however we observe that from physical considerations the

value of $\phi(T)$ must oscillate about 1 *i.e.* the non dimensional equilibrium radius. It must

be noted here that the Villermaux and Bossa (2009) analysis suggests that the drop

oscillates about a mean corresponding to a drop radius of zero which seems inconsistent

and shall be resolved in the discussion that ensues. In view of the above arguments and



following Plesset (1977) we seek an expansion for the non-dimensional radius, $\phi(T)$ as shown below,

$$\phi(T) \sim \phi_0(T) + \delta\phi_1(T) + O(\delta^2) \text{ where, } |\delta| < 1. \tag{4.14}$$

$\phi_0(T)$ is the equilibrium state of the drop and the subsequent terms are higher order corrections in $\varepsilon$ to this non dimensional radius. Substituting (4.14) in (4.12) we obtain,

$$\delta^0: \quad \ddot{\phi}_0 + \frac{Oh}{\sqrt{We}} \frac{\dot{\phi}_0}{\phi_0^2} - 4\left(\frac{a^2}{16} - \frac{6}{We}\right)\phi_0 = 0. \tag{4.15}$$

$$\delta^1: \quad \ddot{\phi}_1 + \frac{8Oh}{\sqrt{We}}\left(\frac{\dot{\phi}_1}{\phi_0^2} - 2\frac{\phi_1\dot{\phi}_0}{\phi_0^3}\right) - 4\left(\frac{a^2}{16} - \frac{6}{We}\right)\phi_1 = 0. \tag{4.16}$$

In (4.14) $O(\delta^2)$ terms are neglected , $O(\delta^0)$ terms correspond to the equilibrium position of the drop, and $O(\delta^1)$ terms represent the deviations about this equilibrium state. The balance between the surface tension and aerodynamics disrupting forces exists in this equilibrium state and the resulting expression for the diameter of the drop is,

$$d_0 \sim \frac{\sigma}{\rho_g U^2}. \tag{4.17}$$

(4.12) now transforms into a linear second order differential. Using the conditions, $\dot{\phi}_0\left(T\right) = 0$ and $\phi_0\left(T\right) = 1$ which correspond to a sphere in the equilibrium state we arrive at the more familiar equation for a harmonic oscillator with damping.

$$\ddot{\phi}_1 + \frac{8Oh}{\sqrt{We}}\dot{\phi}_1 + |k_r|\phi_1 = 0. \tag{4.18}$$



(4.18) is similar to that representing a linear spring mass damper system with the restoring force given by the coefficient of $\phi_1$ and the damping term by the coefficient of $\dot{\phi_1}$.

The stiffness $k_r = k_r\left(a, We\right)$ is given by,

$$k_r\left(a, We\right) = \begin{cases} \left(\dfrac{24}{We} - \dfrac{a^2}{4}\right) & We < \dfrac{96}{a^2} \\ 0 & We = \dfrac{96}{a^2} \\ -\left(\dfrac{24}{We} - \dfrac{a^2}{4}\right) & We > \dfrac{96}{a^2} \end{cases}. \tag{4.19}$$

The stiffness term (4.19) takes different values ranging from negative to positive depending upon $We$. In accordance with the nature of the restoring force we ensure that this value is positive which justifies the use of $|\;|$ with $k_r$. Intuitively, with increasing $Oh$ we expect the value of the critical $We$ number at transition to increase. Hence, to derive the condition for criticality we use $We > \dfrac{96}{a^2}$. Such a calculation considers that the drop oscillates even at $We < \dfrac{96}{a^2}$ which is true within $\dfrac{96}{a^2}\left(1 - \dfrac{2}{3}Oh^2\right)$ of $\dfrac{96}{a^2}$. This seems to be an artifact of the term $|k_r|$ of (4.19) which introduces the anomaly of growing solutions for $We < \dfrac{96}{a^2}$ for $Oh > 0$. As per Hsiang and Faeth (1995), oscillations in the drop cease beyond $Oh = 0.3$ which implies that the growth of $\phi_1(t)$ exists from $0.94\left(\dfrac{96}{a^2}\right)$ to $\left(\dfrac{96}{a^2}\right)$ i.e. for 6% of $\left(\dfrac{96}{a^2}\right)$.



Reverting to (4.18) and setting the condition for the solutions not to grow exponentially, we obtain (4.20),

$$We = We_{c\,Oh\to 0}\left(1 + \frac{2}{3}Oh^2\right).$$  (4.20)

$We_{c\,Oh\to 0}$ is the critical $We$ at $Oh = 0$ (or the inviscid case). In (4.12) for the limit of $Oh = 0$ we obtain $We_{c\,Oh\to 0} = \frac{96}{a^2}$. We choose a value of $a$ equal to $2\sqrt{2}$, comes from the PIV measurements of Flock *et al.* (2012) who give an approximate value of $a$ as 3.0 for ethanol drops. This leads to a value for $We_{c\,Oh\to 0}$ of 12 and is in good agreement with the value reported in reviews of Pilch and Erdman (1987) and Guildenbecher (2009).

To the authors' knowledge this is the first theoretical model which does not involve use of any experimental correlations, such as those seen in the works of Tarnogrodski (1993, 2001), Cohen (1994) and Zhao (2010, 2011). The only constant that needs to be ascertained beforehand is $a$. Furthermore, good match is found between the experimental and theoretical values validating equation (4.20). (4.18) also gives us quantitative information on the frequency of the damped vibrations associated with viscous drops. This value turns out to be $\sqrt{\frac{1}{\tau^2}\left(\frac{24}{We} - 2\right) - \frac{8Oh^2}{We}}$. Setting $Oh = 0$, we get the frequency of oscillations as $\tau^{-1}\sqrt{\frac{24}{We} - 2}$ as expected for inviscid drops.

It is also noteworthy that for $Oh = 0$ (4.15) and (4.16) are decoupled *i.e.* $\phi_0$ does not appear in (4.16). Thus, the expression for the equilibrium diameter remains the same even



for the inviscid case, and (4.16) now reflects the oscillations of the inviscid drop about the equilibrium diameter as given by (4.17). The Villermaux and Bossa (2009) expression can be better understood in the light of the above arguments.

### 4.4 Thickness of the Oblate Spheroid, $h$ ($t$)

From (4.6) we obtain,

$$\frac{dH}{dT} + \frac{2H}{\phi}\frac{d\phi}{dT} = 0. \qquad (4.21)$$

In (4.21) $h(t)$ is non dimensionalised by $d_0/2$

$$H \sim \phi^{-2}. \qquad (4.22)$$

Solving (4.21) for the inviscid case *i.e.* $Oh = 0$ we obtain the expression,

$$\phi \sim e^{\sqrt{m}T}. \qquad (4.23)$$

where, $m = \left(2 - \dfrac{24}{We}\right)$ taking the value of $a$ as $2\sqrt{2}$. $\phi = cosh\left(\sqrt{m}T\right)$ in an exact sense but

(4.23) works well as an asymptotic solution for sufficiently large $T$.

Another relevant parameter in the atomization context that we can extract from the given equations is $T_{ini}$, the initiation time. Using (4.23) we can compute the initiation time for inviscid drops as written below,

$$T_{ini} = \frac{cosh^{-1}\phi_{max}}{\sqrt{2 - \dfrac{24}{We}}}. \qquad (4.24)$$



(4.24) depends on $\phi_{max}$, which is the value of $\phi$ just before bag formation. It is not possible to estimate this from (4.21) as it predicts an indefinite growth of $\phi$ and does not indicate when the bag formation will begin. One must note that $h(t)$ does not shrink indefinitely in time as the rim eventually destabilizes. We, however, attempt to compare (4.24) with existing correlations. Dai and Faeth (2001) give a value of $\phi_{max}$ as 2.15 for 20 $< We < 80$ and $Oh < 0.1$. Substituting this in (4.24) $T_{ini}$ transforms to $1.4\left(2 - \dfrac{24}{We}\right)^{-\frac{1}{2}}$ giving a mean value of 1.34, which is somewhat close to the value of 1.5 quoted in the Guildebecher *et. al.* (2009) review.

## 4.5    Bag Growth, $\beta(T)$

Bag growth plays an important role in the drop disintegration process. As the drop deforms and extends along the cross stream direction (while contracting in the streamwise direction) there is an unequal pressure distribution, $\Delta p$, around the drop. This process ends with a gradual outward bulging of the oblate spheroidal structure, which happens to accommodate the growing $\Delta p$ across the drop. In this section we attempt to model the bag shape and establish its relationship with *We* and *Oh*.

Writing the following force balance at the bag tip and making *We* corrections to the Villermaux and Bossa (2009) expression by incorporating the surface tension term, we obtain,



$$\ddot{\beta} - \frac{24}{We}\beta - 2e^{T\sqrt{4-\frac{96}{We}}} = 0. \qquad (4.25)$$

Here, $\beta = \beta(T) = \frac{\alpha(T)}{d_0/2}$ where $\alpha(T)$ is shown in Chapter 3. The initial conditions for this differential equation are $\dot{\beta}(0) = 0$ and $\beta(0) = 0$. The surface tension term involves evaluating the curvature term given by $\frac{2}{r_c}$, where $r_c$ varies with every point on the bag. However, since we are concerned with the dynamics near the bag tip this variation can be ignored. From geometrical considerations shown in Fig. 4.1 we observe that the radius of curvature, $r_c$, at the bag tip is,

$$r_c = \frac{R^2(t) + \alpha^2(t)}{2\alpha(t)}. \qquad (4.26)$$

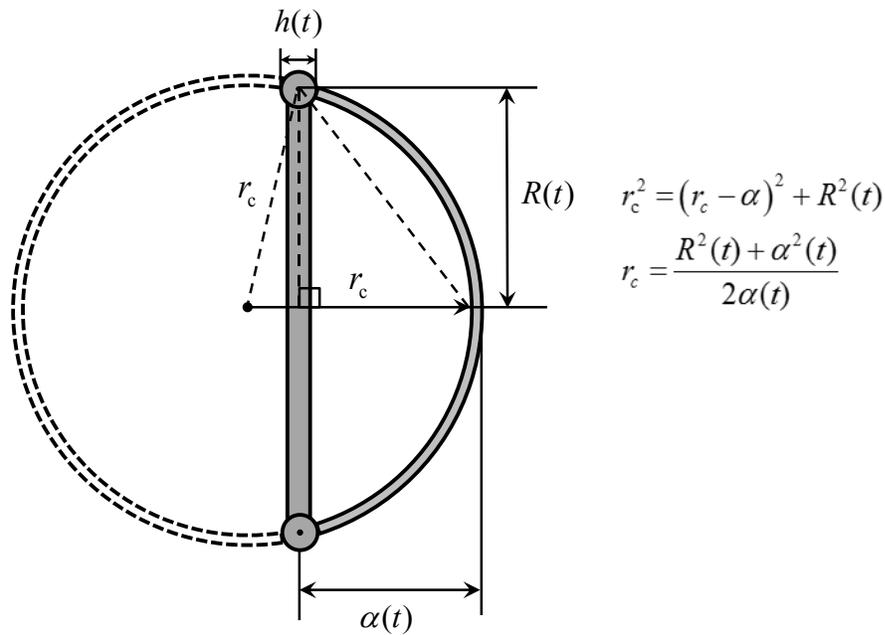

Figure 4.1: Sketch of the expanding bag showing the relevant parameters.



For simplification we assume, $\dfrac{R^2(t)}{2\alpha(t)} > \dfrac{\alpha(t)}{2}$, which is true for bag breakup. Using (4.26) in the force balance at the bag tip we obtain (4.25).

From (4.25) we observe bag growth which increases with increasing *We* at a given instant of time *T*. It is worthwhile to note that (4.25) includes the effect of *We*, a fact not taken into consideration by Villermaux and Bossa (2009) even for the inviscid case.

Employing viscous corrections in (4.25) and evaluating the viscous stress term $2\mu\dfrac{\partial u_{r_c}}{\partial r_c}$ with $u_{r_c} = \dfrac{r_c^2}{R^2}\dot{R}$ for the expanding bag (assuming spherical geometry) results in (4.27).

$$\ddot{\beta} - \frac{24}{We}\beta + \frac{8Oh}{\sqrt{We}}\left(\phi\dot{\phi} - \frac{\dot{\phi}^2\dot{\beta}}{\beta}\right) - 2\phi^2 = 0. \qquad (4.27)$$

The same non-dimensionalization as used in (4.12) is invoked here. As we can readily observe that (4.27) requires the value of $\phi(T)$, which can be obtained from (4.22), but that corresponds to the case of an inviscid drop. However, we need $\phi(T)$ for a viscous drop, this can be obtained by solving (4.12) analytically. Since this is not possible we circumvent these manipulations by using (4.22) in (4.27). This is justified from the Hsiang and Faeth (1995) experimental data where we see that for $Oh > 0.3$ the oscillatory mode ceases to exist, and a 7% change in $\phi(T)$ is seen between $Oh = 0$ and $Oh = 0.3$ at $T = 2$, which roughly corresponds to the end of the bag expansion process. (4.28) is thus arrived at and includes the effect of viscosity and surface tension,



$$\overset{\bullet\bullet}{\beta} - \frac{24}{We}\beta + \left[\frac{8Oh}{\sqrt{We}}\sqrt{\left(2 - \frac{24}{We}\right)} - 2\right]e^{T\sqrt{4 - \frac{96}{We}}} = 0. \qquad (4.28)$$

Setting $Oh = 0$ we recover (4.25). To capture the dynamics at other points on the bag one can follow the arguments of Joseph *et.al.* (2003) for a rising spherical cap and construct at a given time $T$, $\beta(\theta, T) = \beta(0, T)(1 + s\theta^2)$ with $\theta = 0$ at the bag tip and $s = \frac{1}{\beta(0, T)}\frac{\partial^2 \beta}{\partial \theta^2}\bigg|_{\theta=0}$ ($s$ is the measure of deviation from perfect sphericity which exists near the stagnation point). This is redundant to the current analysis, and hence not dealt with any more detail here.



CHAPTER 5.  RESULTS AND DISCUSSION

This chapter presents representative model predictions and compares them to experimental data. Salient breakup related parameters such as initiation time, $T_{ini}$, $We$ v/s $Oh$ scaling at breakup regime boundaries, bag evolution, and drop sizes will be reported. To conclude, an account of drop size measurements is given.

## 5.1    Inviscid and Viscous Drops

Three comparisons between experimental and theoretical findings are presented. The first is the $We$ versus $Oh$ scaling at the breakup regime boundaries. The second is the initiation time. The third is bag sizes at various times. Towards the end, a discussion on the fragment sizes and various instabilities involved is presented.

Fig. 5.1 shows a typical high speed video image sequence demonstrating the various stages of bag breakup for inviscid and Newtonian drops. They are as outlined in Chapter 1.



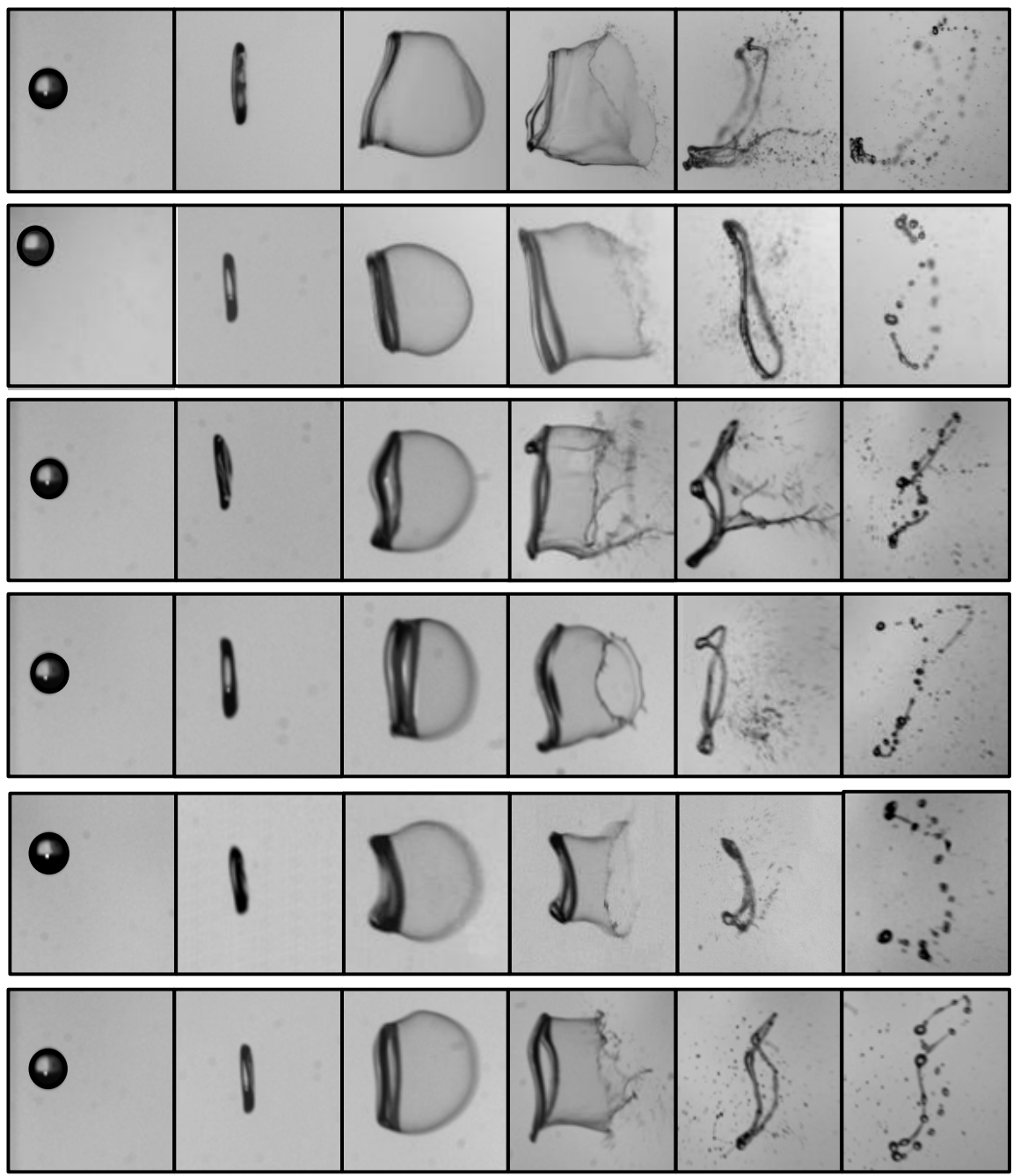

**4 mm**

Figure 5.1: Images showing bag breakup of drops. Top to bottom: *Oh* = 0.002, 0.007, 0.010, 0.015, 0.034, 0.058 for *We* ~ 14. Δ*t* between successive frames is roughly 2 *ms*.



### 5.1.1  *We* v/s *Oh*

Predictions using the Chapter 4 model show the variation of *We* with *Oh* is similar to that obtained in the empirical relation of Pilch and Erdman (1987). Our predictions as obtained in equation (4.20) show the exponent of *Oh* to be 2, which is comparable to the value of 1.6 obtained by Brodkey (1987) when he correlated the existing experimental data. The comparisons are shown below. $We_{cr} = 12$ was taken for the sake of comparison. For values of *Oh* < 1, the predicted and experimentally observed values show a good match.

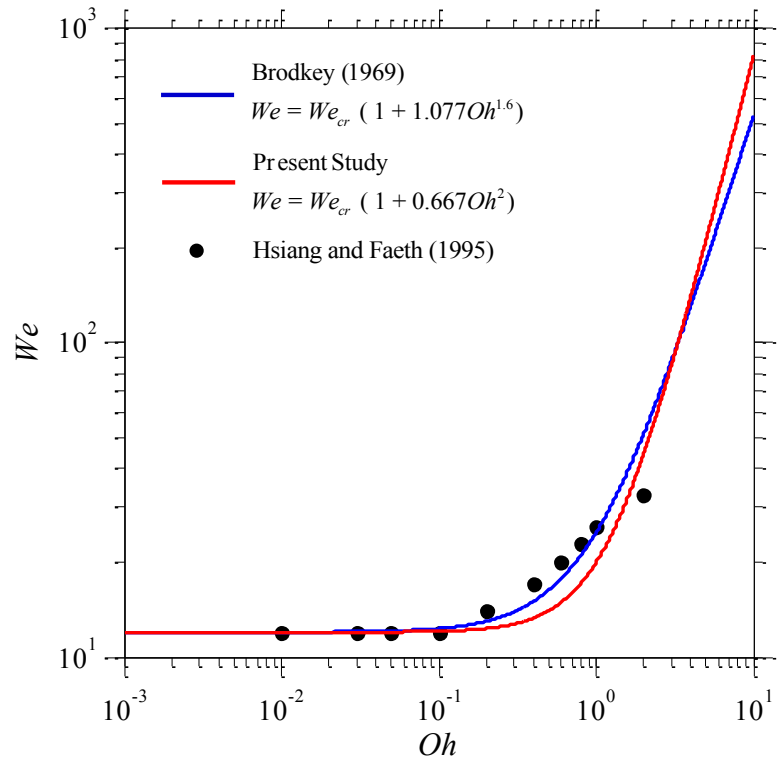

Figure 5.2: *We* v/s *Oh* plot indicating onset of bag breakup.



### 5.1.2   Radial Growth, $\phi(T)$

Equation (4.12) gives a measure of the radial growth of the deforming drop as it transits from a sphere to an oblate spheroid. This non-linear ODE is solved using the ode45 solver of MATLAB by reducing the second order ODE to a system of two equations. The radial growth is represented by $\phi(T)$ and plotted in Fig. 5.3. We can see an exponential growth in the radial extent until we reach the oblate spheroid stage, after which the bag forms.

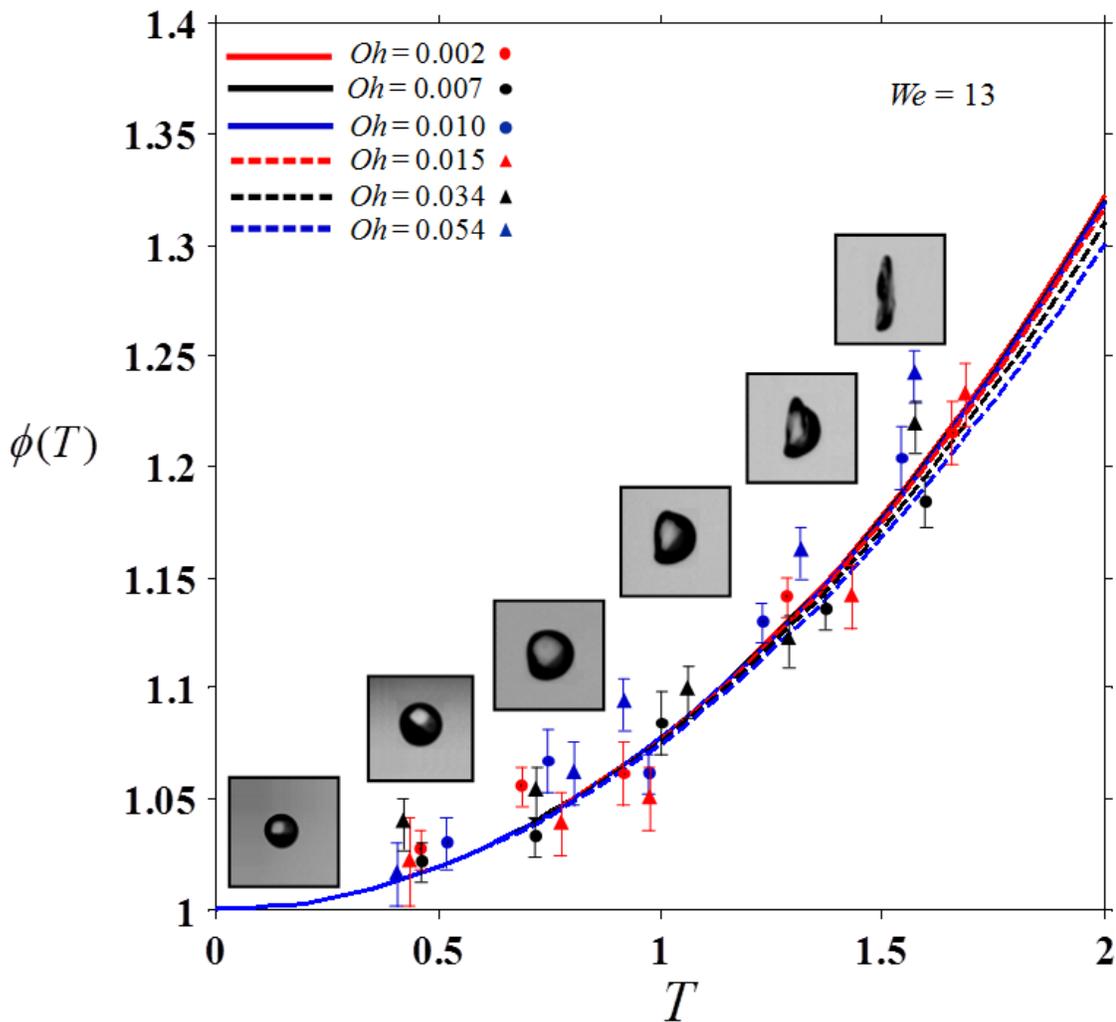

Figure 5.3: Variation of $\phi(T)$ *v/s* $T$ for a fixed *We* (=13). Symbols are experimental data while lines represent theoretical results.



The effect of *Oh* for a given *We* is studied. For the range of permissible *Oh* we do not see a predominant effect of this variation.

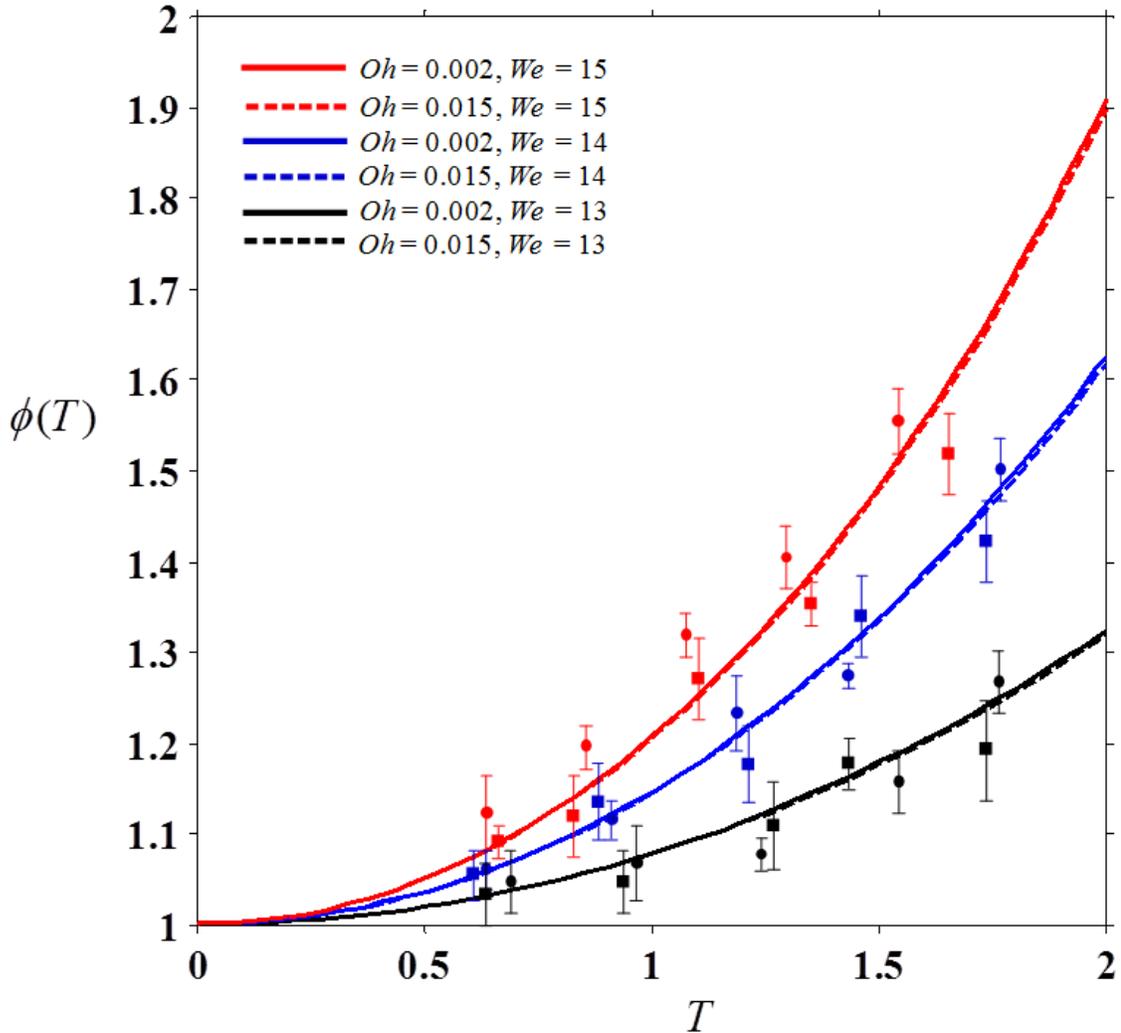

Figure 5.4: Variation of $\phi(T)$ *v/s T* for varying *We* and *Oh* = 0.002, 0.015. Symbols are experimental data while lines represent theoretical results.

In Fig. 5.4 we see that *We* does have a dominant effect on the radial growth of the deforming drop. This is true at even high *Oh* as the aerodynamic disruptive forces are higher than the restoring surface tension forces.



### 5.1.3    Validation of the Velocity Field, $u_r(r,t)$

The velocity field [equation (4.8)] obtained as a solution to the equation for the conservation of mass is verified experimentally in Fig. 5.5. Using the expression for $\phi(T) = \sqrt{2 - \dfrac{24}{We}} \, cosh(T)$ we get the non-dimensional version of the velocity field $U(T)$ [= $\dot{\phi}(T)$]. It agrees well with the experimental findings and validates the equation.

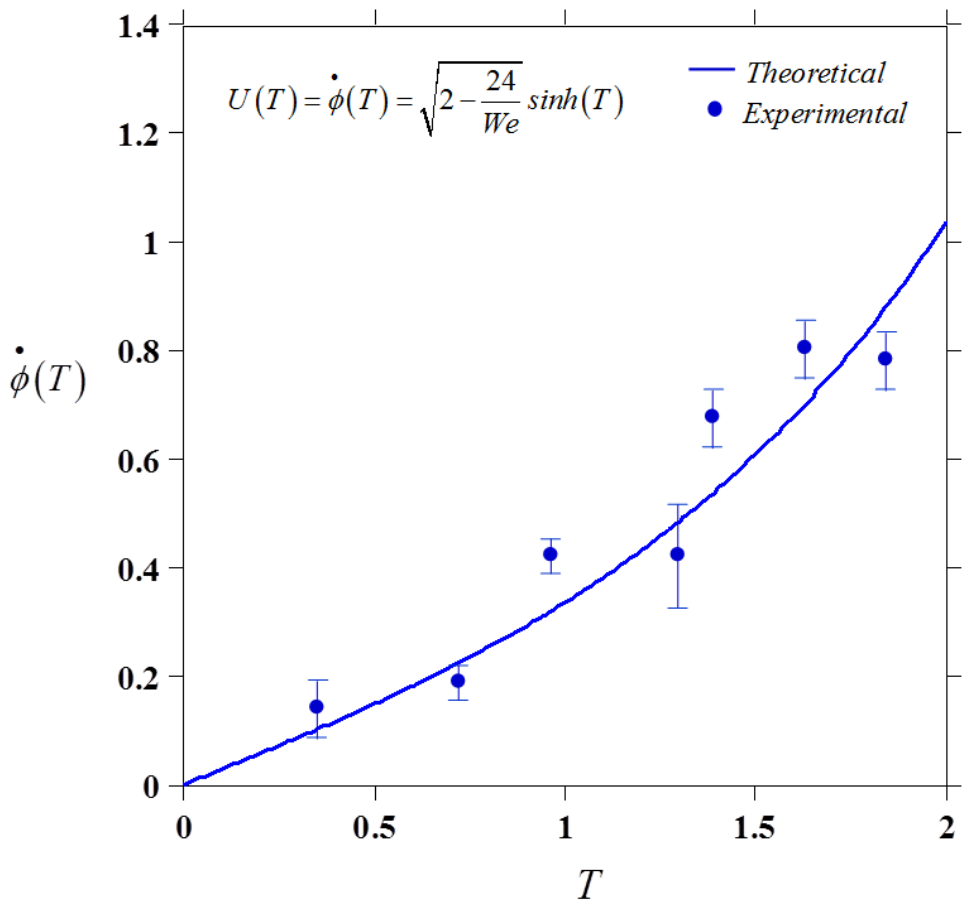

Figure 5.5: Liquid flow field velocity, $U(T) = \dfrac{u_r(r,t)\big|_{r=R}}{d_0 \big/ \tau\big/_2}$.



### 5.1.4   Initiation Time ($T_{ini}$)

Fig. 5.6 shows comparison between $T_{ini}$ predictions [equation (4.24)] and experimental measurements. The mean of 1.34, as obtained from theory compares satisfactorily to the value of 1.5 as reported in the Guildenbecher *et al.* (2009) review. One must bear in mind that determining this value experimentally is subject to discretion.

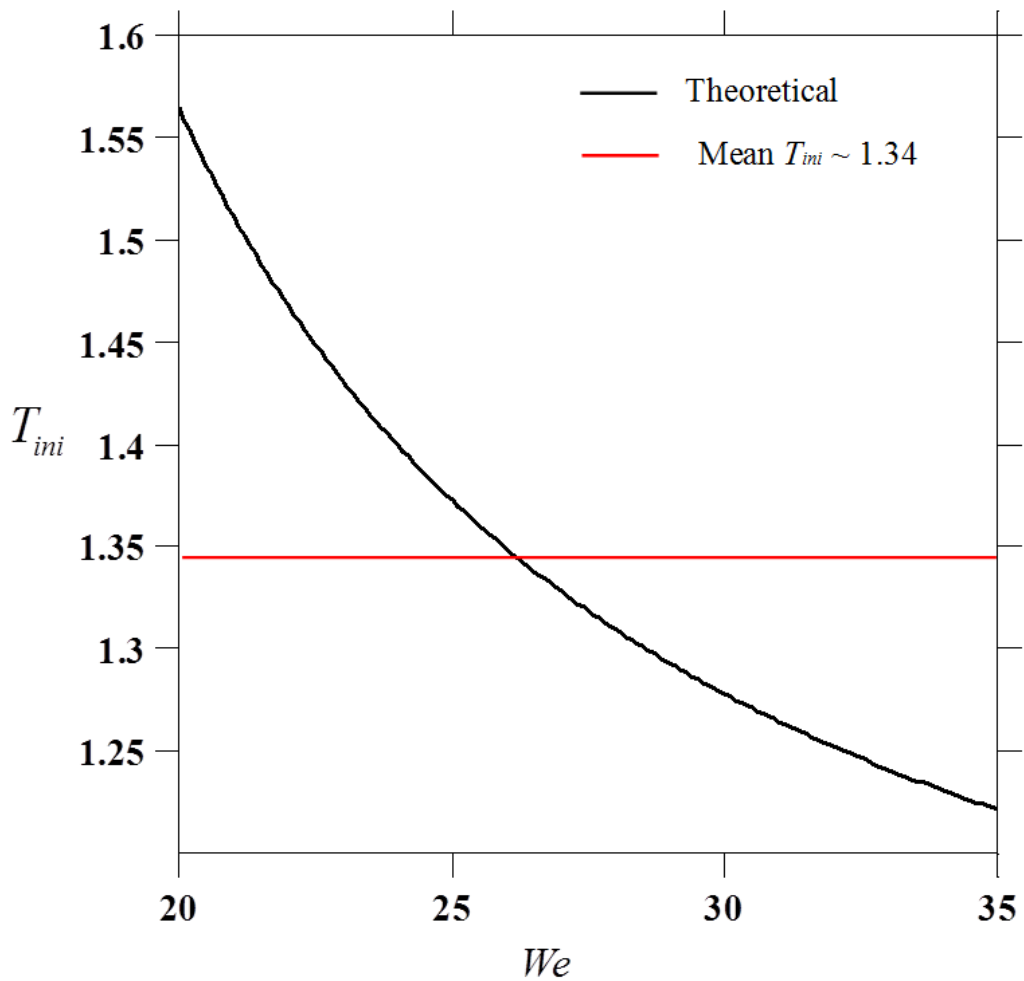

Figure 5.6: *$T_{ini}$ v/s We.*



### 5.1.5   Bag Growth, $\beta(T)$

The growth of the bag is theoretically predicted using (4.25) for the inviscid case incorporating *We* corrections. The bag growth is denoted by non-dimensional $\beta(T)$. Two comparisons are made: (*i*) to study the effect of *We* for a given *Oh* (Fig. 5.7) and (*ii*) to study the effect of *Oh* for a given *We* (Fig. 5.8). In (*i*) the effect of *Oh* is insignificant for a given *We*.  Contrary to this observation we see that in (*ii*) we observe a pronounced influence of *We* on the bag expansion extent. Even small changes in *We* alter the bag growth drastically.

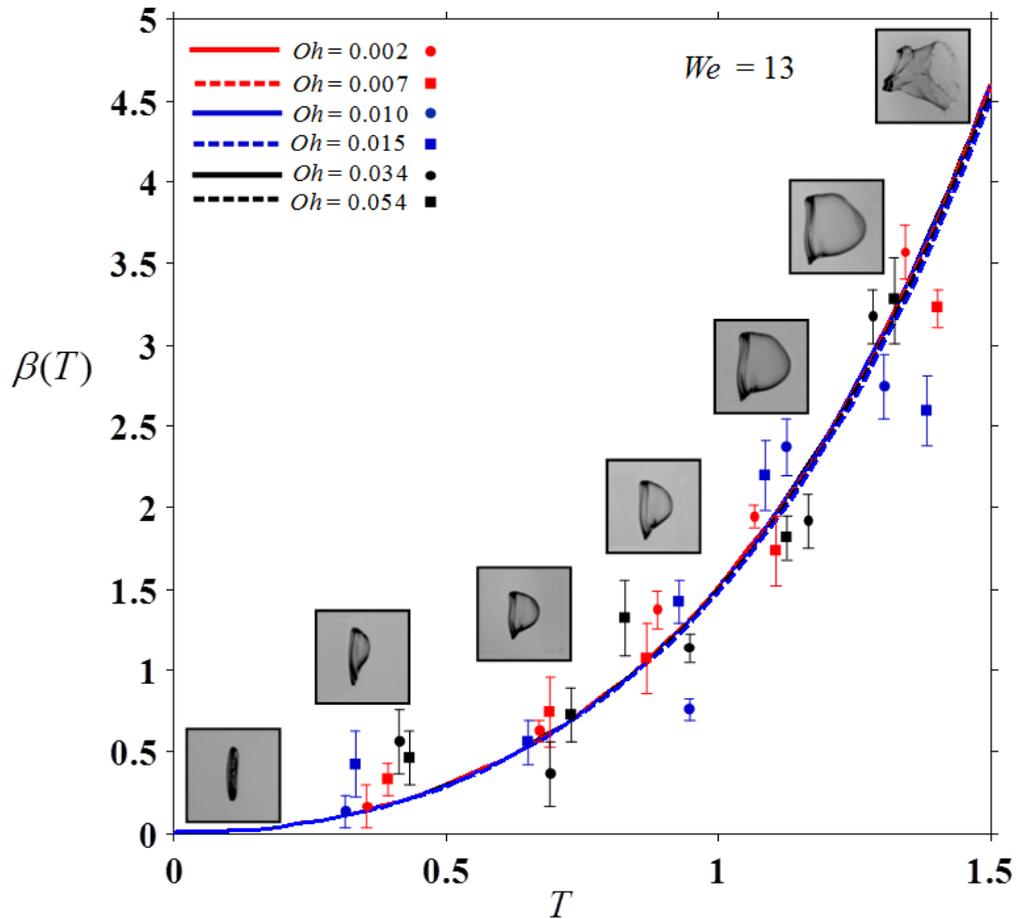

Figure 5.7: Variation of $\beta(T)$ v/s $T$ for varying *Oh* and given *We*. Symbols represent experimental data while lines represent theoretical results.



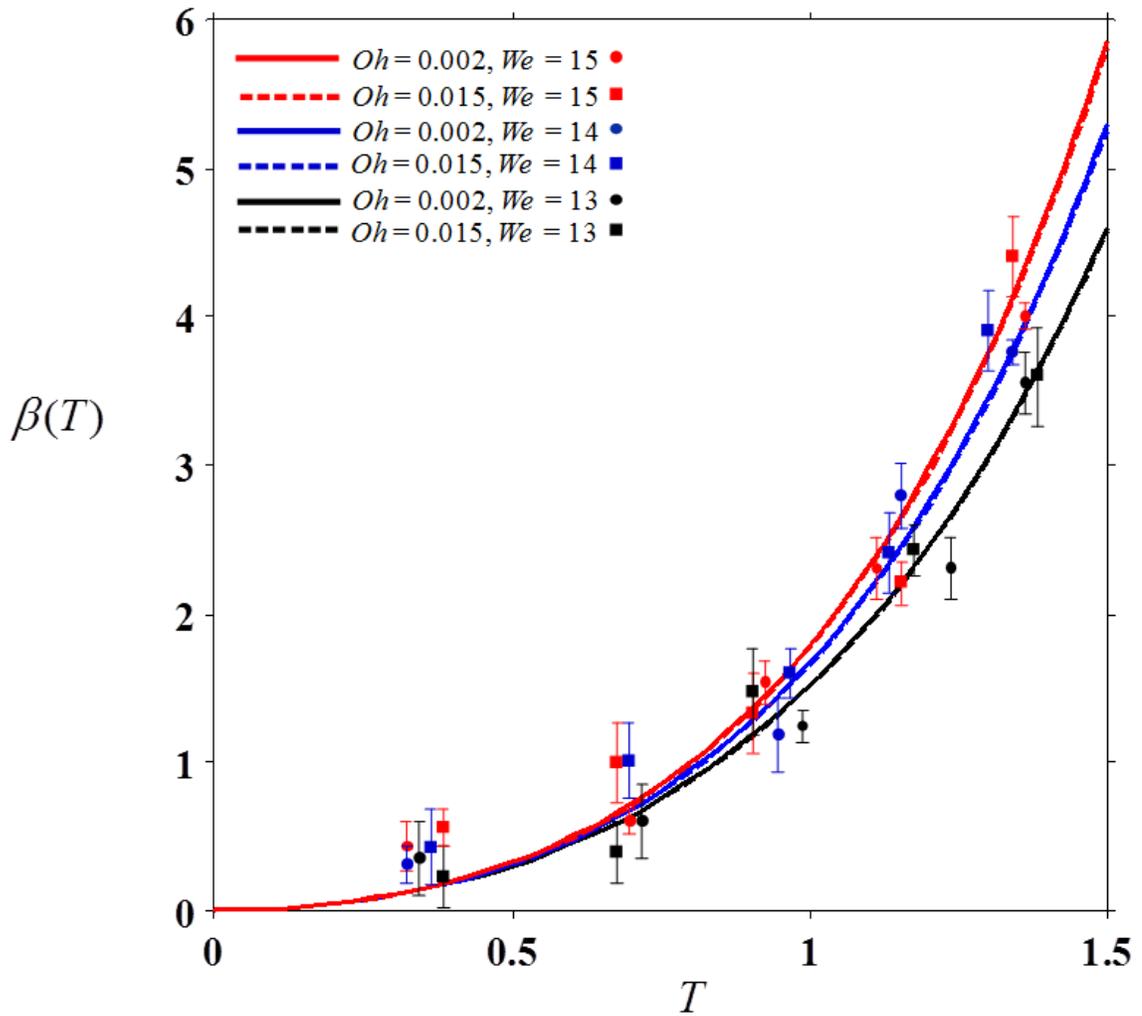

Figure 5.8: Variation of $\beta(T)$ v/s $T$ for varying $We$ (= 13−15) and $Oh$ = 0.002, 0.015. Symbols represent experimental data while lines represent theoretical results.



### 5.1.6    Drop Size Distribution

Following the bag expansion, the deformed drop begins to show signs of disintegration. This is marked by collapse of the two main topological features of the deformed drop (*i.e.,* the bag and the rim). However, they do not collapse simultaneously with the bag rupture preceding the rim disintegration. The bag grows exponentially, and as noted by Villermaux and Bossa (2009), no finite time singularity is observed for such deformation under the influence of normal stresses. As expected, the bag and the rim fragments constitute the drop ensemble. We expect drops of two characteristic sizes, one corresponding to the rim and the other to the bag. This ideally should lead to a bi-modal distribution. However, this is not seen in our measurements (Fig. 5.9, Fig. 5.10) as the rim fragments are small in number and are overwhelmingly outnumbered by the bag fragments. For a given *We,* Fig. 5.9 shows that even for the small values of *Oh* that we have tested a drastic drop in the peak is seen. This can be attributed to more even-sized fragments as a consequence of higher bag thickness due to higher viscosity. On the other hand, in Fig. 5.10 a slight drop in the peak is seen for a given *Oh* and decreasing *We,* which is an indicator of a large number of drops of smaller size because of higher fragmentation. Fig. 5.11 gives an estimate of the non-dimensionalised Sauter mean diameter, (SMD or $D_{32}$) which lies between 0.036 and 0.055. This is consistent with the values predicted by Dai and Faeth (2001), who actually get a higher value from their measurement as they primarily measured the rim fragments.



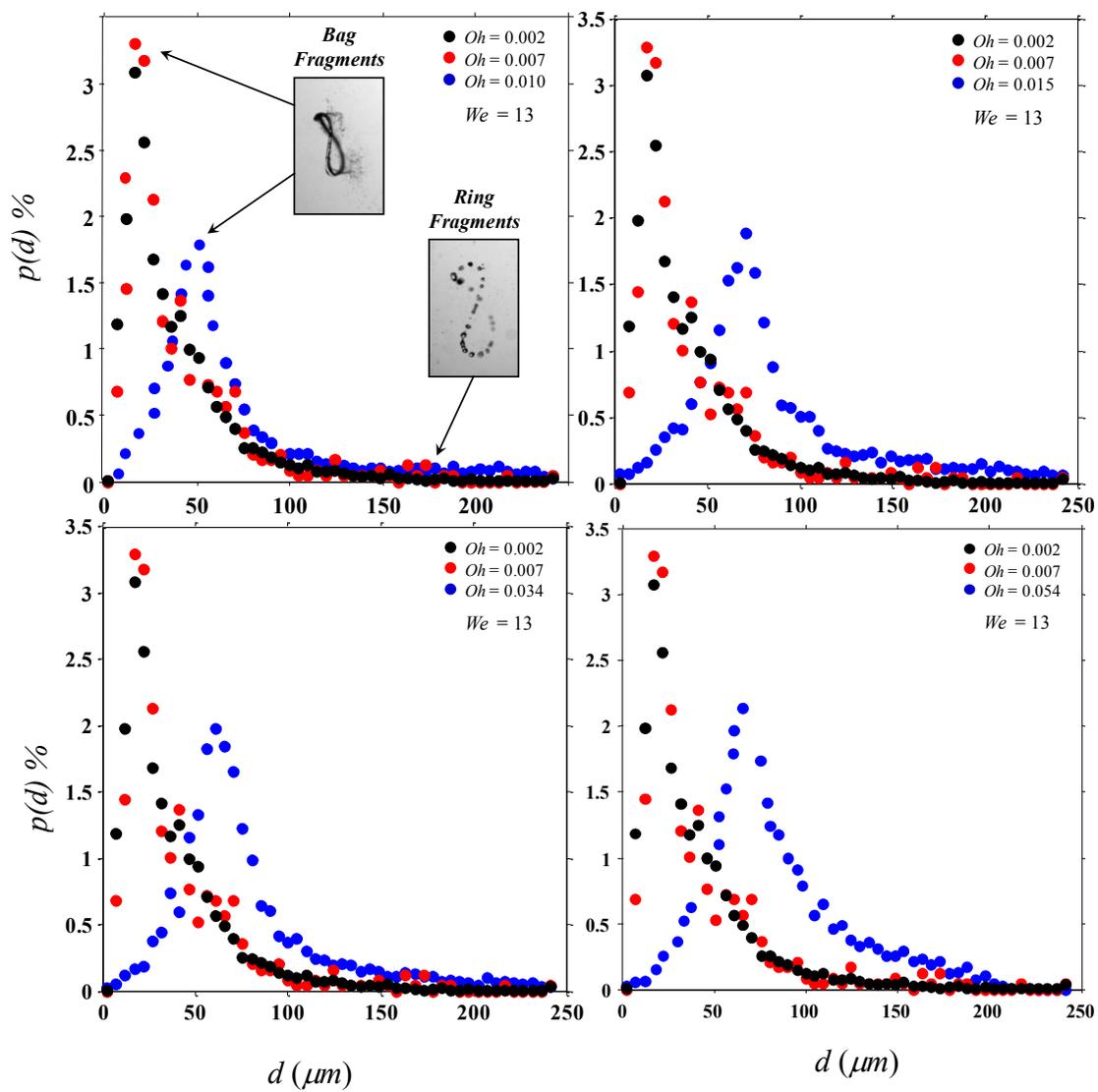

Figure 5.9: *pdf v/s d* for constant *We* = 13 and varying *Oh* = 0.010, 0.015, 0.034, 0.0054.



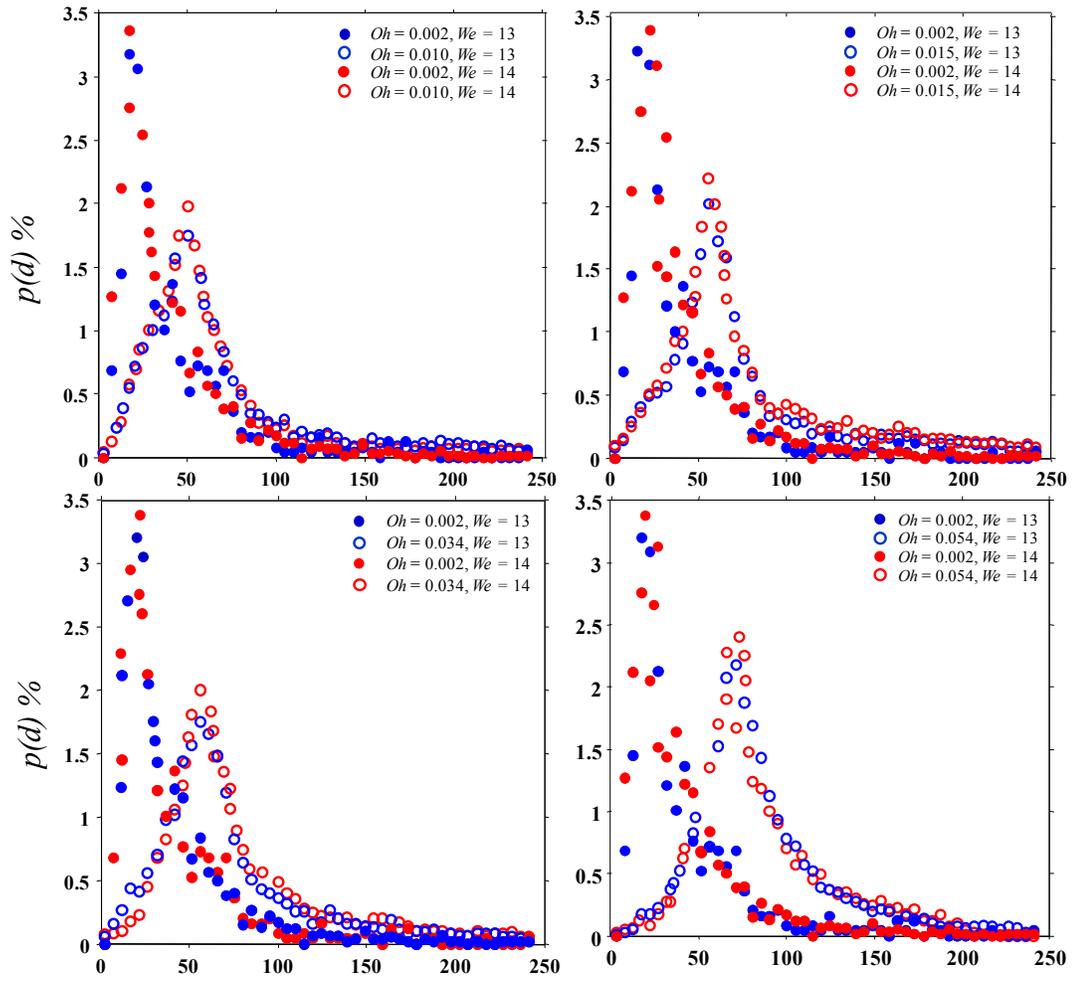

Figure 5.10: *pdf v/s d* for *We* = 13, 14 and *Oh* = 0.002, 0.010; 0.002, 0.015; 0.002, 0.034; 0.002, 0.054.



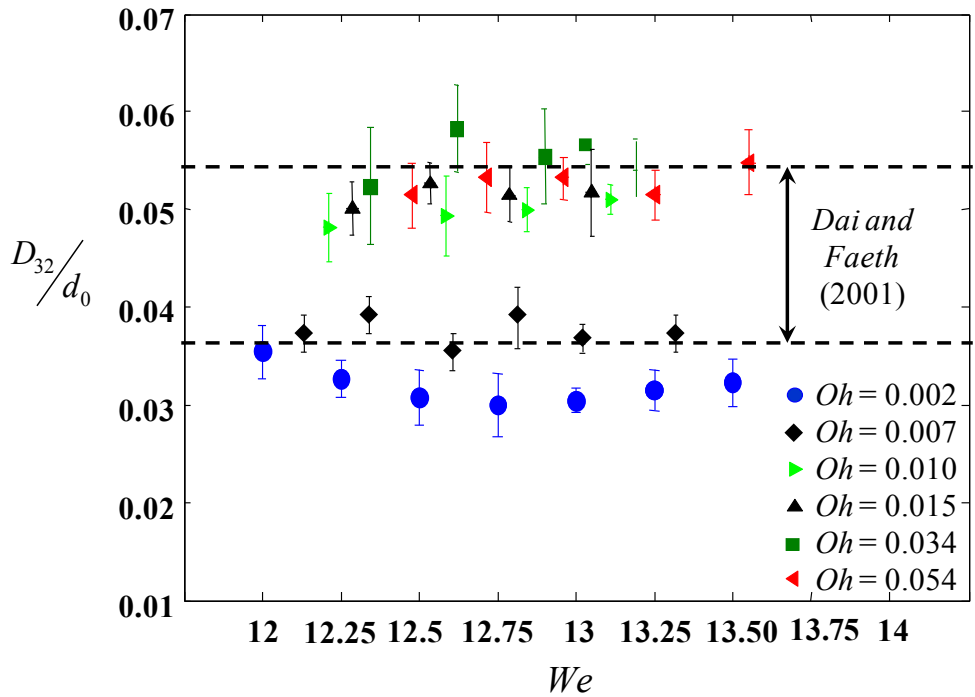

Figure 5.11: Variation of SMD ($D_{32}$) with $We$ for varying $Oh$.

The mechanisms governing rim fragmentation can be explained based on Plateau–Rayleigh instability or Rayleigh–Taylor instability. The role of Rayleigh–Taylor instability although may appear less likely in such a scenario (Lhussier and Villermaux, 2011; Bush and Hasha, 2004; Kretchcnikov, 2010), but it cannot be readily eliminated and will be carefully evaluated. For our experimental solutions, the viscous characteristic time scale, $\tau_{vis} = \ell \mu_l \big/ \sigma$ is small compared to the inertial time scale, $\tau_{in} = \sqrt{\rho_l \ell^3 \big/ \sigma}$ indicating that the growth of disturbances is driven by interplay between inertia and surface tension. Here, $\ell$ is the length scale, which is the rim thickness, $h(t)$, or the bag thickness, $t_f$, $\sigma$ is the surface tension of the interface, $\rho_l$ is the density of liquid, and $\mu_l$ is the viscosity of liquid. To keep things general we include the effects of viscosity.



In view of the above statements, we may write the following expressions for maximum growth rate, $\omega_{max_{R-T}}$ and corresponding wavelength, $\lambda_{max_{R-T}}$ for Rayleigh–Taylor instability when the lighter fluid is gas and liquid drop viscosity is taken into account (Aliseda *et al.* 2008).

$$\lambda_{max_{R-T}} = 2\pi\left[\sqrt{\frac{3\sigma}{\gamma\rho_l}} + \sqrt[3]{\frac{\mu_l^2}{\gamma\rho_l^2}}\right]. \tag{5.1}$$

$$\omega_{max_{R-T}} = \left(\frac{4\pi^2}{\lambda_{max_{R-T}}^2\,\rho_l}\right)\left[\sqrt{\mu_l^2 + \frac{\gamma\rho_l^2}{8\pi^3}\lambda_{max_{R-T}}^2 - \frac{\sigma\rho_l}{2\pi}\lambda_{max_{R-T}}} - \mu_l\right]. \tag{5.2}$$

In the above equations $\gamma$ is the acceleration of the liquid interface. For low viscosity fluids like the ones tested here (5.1) and (5.2) reduce to (5.3).

$$\lambda_{max_{R-T}} = 2\pi\left(\frac{3\sigma}{\gamma\rho_l}\right)^{1/2} \qquad \omega_{max_{R-T}} = \left(\frac{4}{27}\frac{\rho_l\gamma^3}{\sigma}\right)^{1/4}. \tag{5.3}$$

For the Plateau–Rayleigh scenario (5.4), (5.5), and (5.6) represent maximum growth rate, most amplified wavelength, and the corresponding expressions for the inviscid case.

$$\omega_{max_{R-P}} = \left[\left(\frac{8\rho_l R_0^3}{\sigma}\right)^{1/2} + \frac{6\mu_l R_0}{\sigma}\right]^{-1}. \tag{5.4}$$

$$\lambda_{max_{R-P}} = \sqrt{8}\,\pi R_0\left[1 + \frac{3\mu_l}{\sqrt{2\rho_l\sigma R_0}}\right]^{1/2}. \tag{5.5}$$

$$\lambda_{max_{R-P}} = \sqrt{8}\pi R_0 \qquad \omega_{max_{R-P}} = \left[\left(\frac{8\rho_l R_0^3}{\sigma}\right)^{1/2}\right]^{-1}. \tag{5.6}$$

$R_0$ is typically the liquid rim diameter.



5.1.6.1 <u>Rim Fragments</u>

As mentioned in earlier sections, the rim breakup can be explained by either Plateau–Rayleigh instability or Rayleigh–Taylor instability. Since the radial extent of the deformed drop is much greater than the thickness of the rim, the straight jet theory of Rayleigh (1878) can be applied here to qualitatively understand the Plateau–Rayleigh instability as found in Zhang *et. al* (2010).

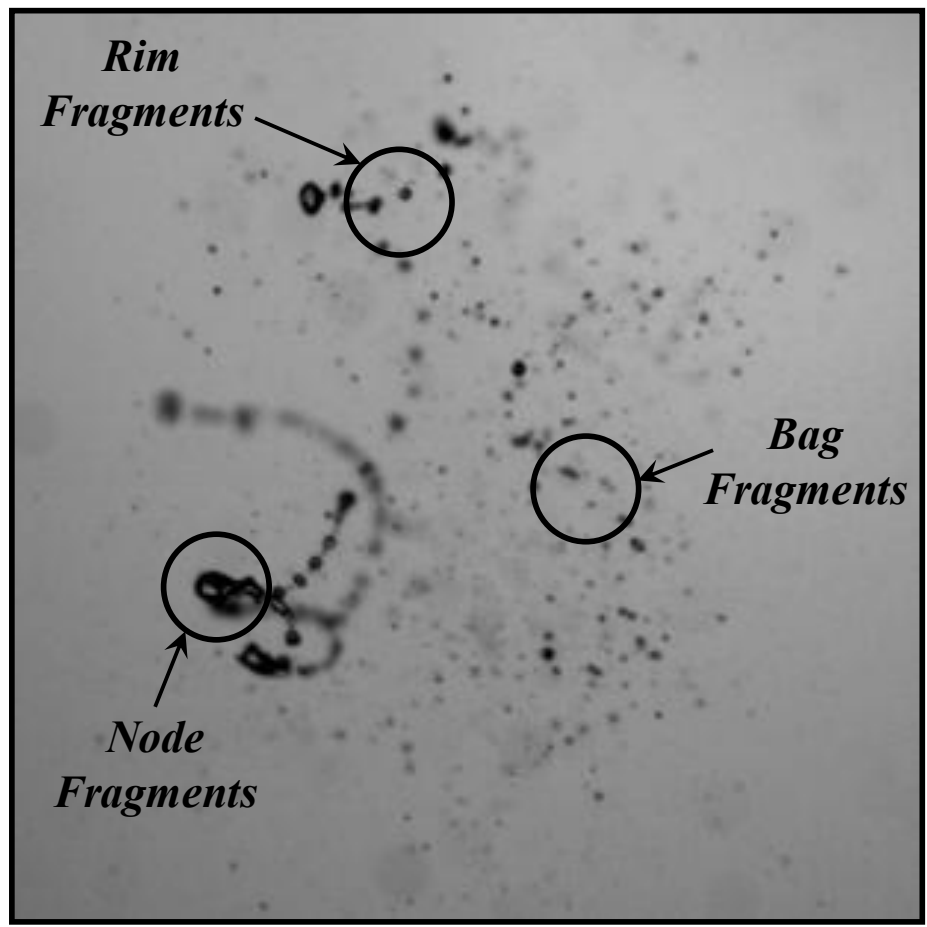

Figure 5.12: Various fragments produced after complete disintegration of the drop.



As seen in our study, the rim is not perfectly smooth. It has visible corrugations on its outer surface and the appearance of nodes (Fig. 5.13). Krechetnikov (2009, 2010) gives an exact treatment of such a problem while dealing with crown splashes generated when a drop impacts a liquid film. Regardless, we assume on the basis of the argument we made to invoke the straight jet theory for Plateau−Rayleigh instability that accounting for the curved edge may be unnecessary. The fragments thus ejected from these nodes are of a different diameter than those which form out of capillary breakup of the rim (Fig. 5.12). This should lead to tri-modality in the drop size distribution. However, we are unable to see this in our experiments as the larger drops produced are very small in number and show up at the tail end of the drop size distribution, such as the one in Fig. 5.9.

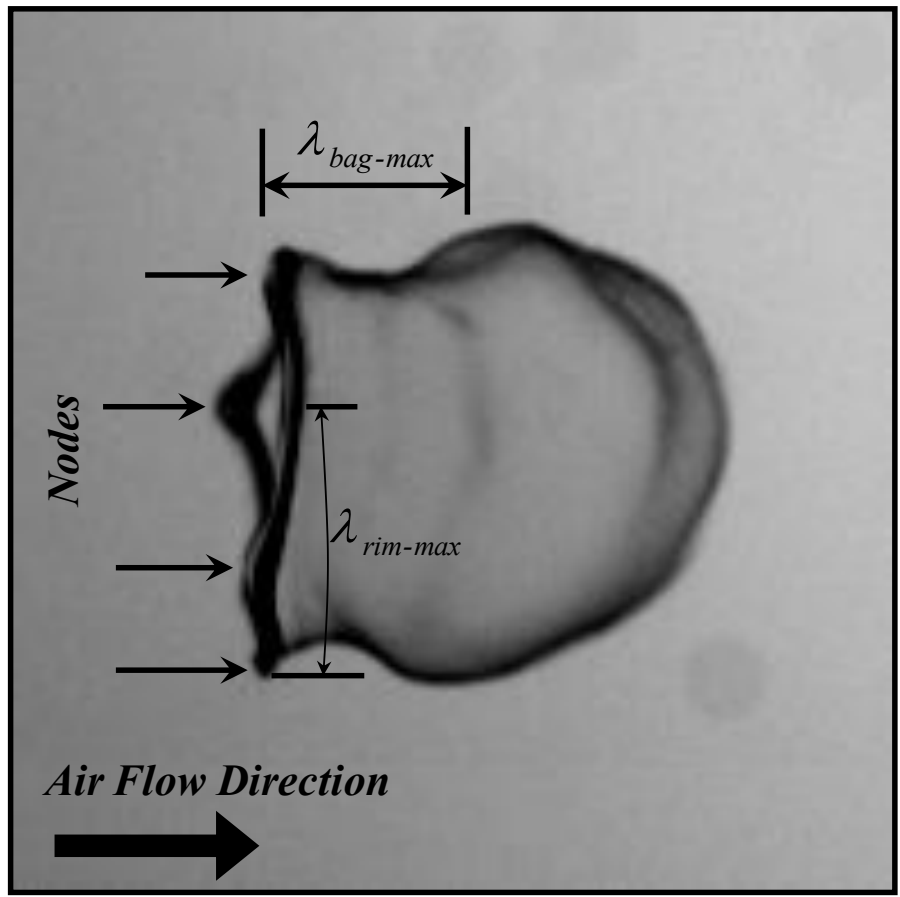

Figure 5.13: Nodes and waves on the drop surface owing to instabilities.



The possibility of the existence of either of these instabilities is a well known fact (Zhang *et al.,* 2010; Roisman, 2010) in the context of thick rims such as the ones we encounter here,

$$\frac{\lambda_{rim\,max_{R-T}}}{\lambda_{rim\,max_{R-P}}} = \frac{1}{\sqrt{2}} \left[ \frac{\left(\frac{3}{Bo}\right)^{\frac{1}{2}} + \left(\frac{Oh^2}{Bo}\right)^{\frac{1}{3}}}{\left(1 + \frac{3}{\sqrt{2}}Oh\right)^{\frac{1}{2}}} \right]^{\frac{1}{2}}.$$ (5.7)

In the above expression, the Bond number *Bo* is,

$$Bo = \frac{\rho_l \gamma \ell^2}{\sigma}.$$ (5.8)

The appropriate length scale, $\ell$, here is the rim thickness $h(t)/2$. Also, the effect of *We* in the denominator of (5.8) is captured via *Oh* which is calculated based on $h(t) = f(We)$ as seen in Chapter 4. For the range of *Bo* (from 10 to 20) and *Oh* (from 0.0001 to 0.01) considered we see that $\frac{\lambda_{rim\,max_{R-T}}}{\lambda_{rim\,max_{R-P}}} < 1$ which means that the predicted maximum Rayleigh–Taylor wavelength is lesser than that due to the Plateau–Rayleigh instability. Fig. 5.14(a) shows that the experimental observations closely match the Plateau–Rayleigh instability predictions instability, shown in Fig. 5.14(b). This implies that the toroidal ring does indeed breakup by the Plateau–Rayleigh mechanism.

Viscosity tends to dampen the growth rate of disturbances and hence the node formation. This pushes their occurrence to higher *We*. Even small *Oh* solutions, as the ones used in this work, are sufficient to damp out their formation at relatively lower *We*.



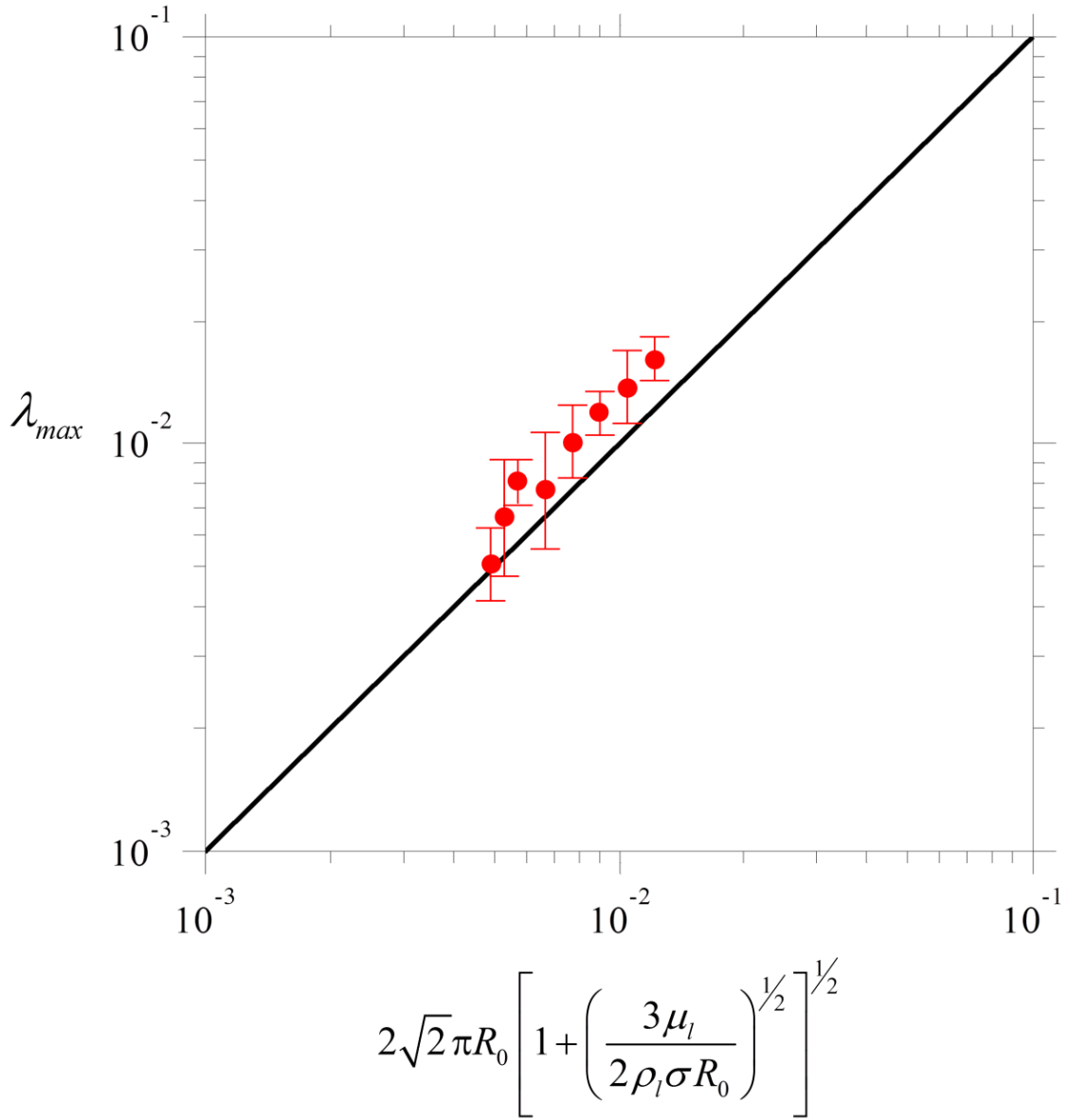

Figure 5.14 (a): Comparison of the experimentally measured wavelength with the Rayleigh−Plateau predictions for the liquid rim.



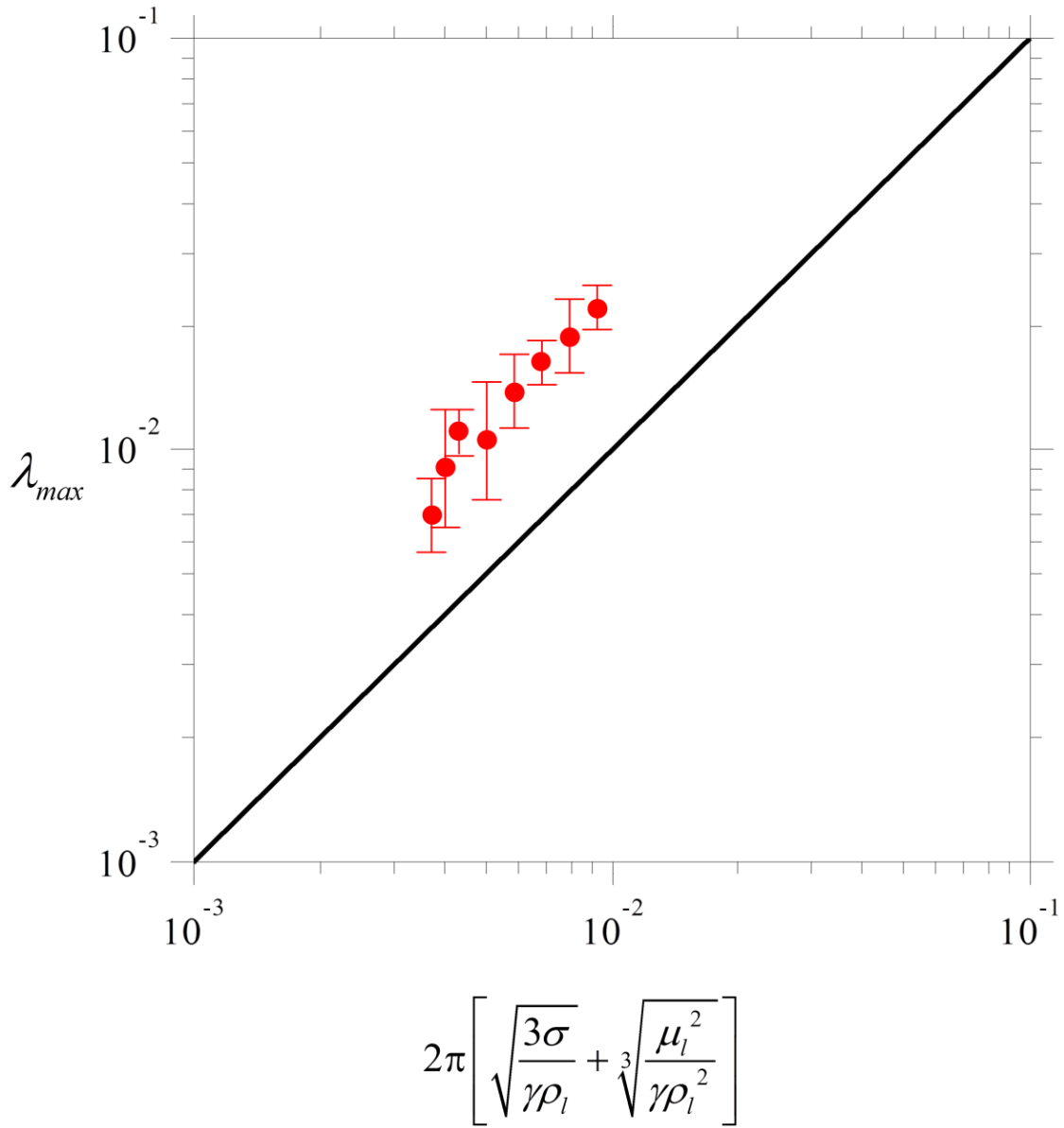

$$2\pi\left[\sqrt{\frac{3\sigma}{\gamma\rho_l}} + \sqrt[3]{\frac{{\mu_l}^2}{\gamma{\rho_l}^2}}\right]$$

Figure 5.14(b): Comparison of the experimentally measured wavelength with the Rayleigh−Taylor predictions for the liquid rim.

### 5.1.6.2 Bag Fragments

The bag expansion is accompanied by changes in bag thickness which become visible on the surface of the liquid film as a result of growing Rayleigh−Taylor instability waves. The expression for critical wavelength $\lambda_{max\text{RT}}$ corresponds to flat sheets as given in (5.2).



Fig. 5.15 shows the comparison between experiments and theory for the bag. We observe a satisfactory match. Significantly, the wavelength corresponding to the liquid rim destabilization is higher than that for the bag.

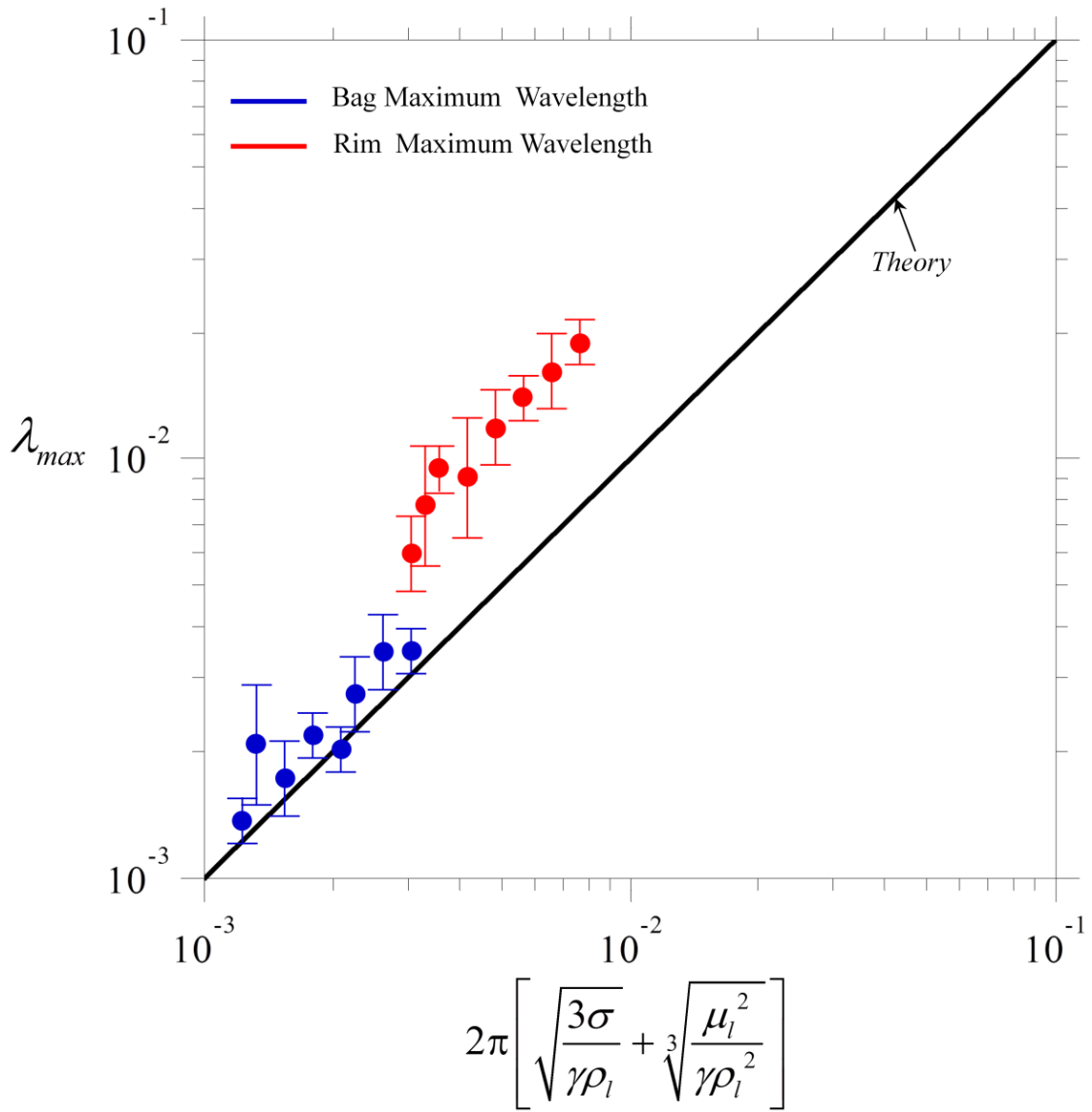

Figure 5.15: Comparison of the experimentally measured wavelength with the Rayleigh-Taylor predictions for the bag and the rim.



### 5.1.6.3 Other Salient Features of the Bag Breakup Process

This growth of Rayleigh-Taylor instability results in the formation of holes on the liquid sheet surface which expand radially with time rupturing the bag as seen in Bremond and Villermaux (2005). Our experiments show that these holes first appear at the tip of the bag where the curvature is minimum. This rate of expansion $V$ of is governed by the Taylor-Cullick law as proposed by Taylor (1959) and Cullick (1960) and applicable for low viscosity films such as those used in this study (Lhussier and Villermaux , 2009). It is given by $V = \sqrt{\dfrac{2\sigma}{\rho_l t_f}}$ where $t_f$ is the bag thickness, $\sigma$ is the surface tension of the interface, and $\rho_l$ is the density of the liquid sheet. Lhussier and Villermaux (2011), Savva and Bush (2009) and Debrage et al. (1995) give details on viscous retracting films.

Towards the closing stages of the breakup event the bursting of the bag occurs, which happens before the rim disintegrates. As pointed out earlier, the rim fragmentation is governed by combined Plateau–Rayleigh/Rayleigh–Taylor instability and the bag only by a Rayleigh–Taylor mechanism. Lhussier and Villermaux (2011) give the characteristic time scale for both of these as,

$$\tau_c \sim \sqrt{\frac{\rho_l \ell^3}{\sigma}}. \tag{5.9}$$

$h(t)\big/2 \gg t_f$ implies $\tau_{bag} \ll \tau_{rim}$, therefore explaining the rupture of the bag prior to the collapse of the ring.



# CHAPTER 6. SUMMARY AND CONCLUSIONS

## 6.1    Summary and Conclusions

The goals of this thesis are threefold. The first is to develop an analytical model that accurately describes the vibrational/bag breakup of inviscid drops. The significant conclusion is that the proposed analytical approach does contain the necessary physical processes.

(*i*) Drop deformation dynamics for various viscous liquid drops were analyzed which included effect of *We* and *Oh*. The radial extent and bag growth are quantified based on these considerations and an exponential growth in both is noted. The effect of increasing *We* is to increase the bag growth and radial extent of the deformed drop at a given time. In contrast, increasing *Oh* dampens this growth, albeit slowly for the solutions considered in this work.

The transition *We* which marks the cross over from vibrational mode to bag breakup is theoretically determined and compared with existing experimental results. The derived expression shows good agreement with these observed experimental values and is an improvement on the relations obtained this far in literature.

(*ii*) A brief account of the initiation time, $T_{\mathrm{ini}}$ is presented and is found to be in good agreement existing literature.

(*iii*) The drop size distribution consisted of two main components: bag and rim fragments the latter being smaller in number hence not visible as a peak in the measured probability distribution. The effect of *Oh* was seen to be significant in reducing the number of smaller



size fragments resulting in lowering of the peak of the distribution. The dependence on *We* is not clearly established, but a slight reduction in the peak of the distribution for low *We*.

(*iv*) An effort is made to bring clarity to the closing stages of bag breakup. The role of Plateau–Rayleigh and Rayleigh–Taylor instabilities is highlighted in rim and bag fragmentation. The Plateau–Rayleigh mechanism is found to explain the rim collapse well and the Rayleigh–Taylor mechanism the bag disintegration. To conclude we analyse the characteristic time scales of the two instabilities and on this basis are able to account for the early bursting of the bag with respect to the rim.

## 6.2    Future Work

The results thus far have been promising. Future investigations are as follows:

• Some aspects pertaining to the hydrodynamic instability analysis can be looked at more closely, such as that for the liquid rim and the bag. At both these places the effect of the curved interface is not included. It may be worthwhile to investigate the role of this factor.

• Extend the viscous Newtonian analytical model to inelastic non-Newtonian liquids. Acquire experimental data necessary for assessing model accuracy.

• Make PDA measurements to determine vibrational/bag breakup mode fragment sizes for inelastic non-Newtonian drops and compare with theoretical analyses.

LIST OF REFRENCES



LIST OF REFRENCES


Aalburg C, Faeth G M, van Leer B (2002), "Deformation and drag properties of round drops subjected to shock–wave disturbances", *AIAA J.*, 41(12): 2371–2378.

Aliseda A, Hopfinger E J, Lasheras J C, Kremer D M, Berchielli A, Connolly E K (2008), "Atomization of viscous and non-newtonian liquids by a coaxial, high-speed gas jet: Experiments and droplet size modeling", *Int. J. Multiphase Flow*, 34: 161–175.

Arcoumanis C, Khezzar L, Whitelaw D S, Warren B C H (1994), "Breakup of Newtonian and non–Newtonian fluids in air jets" , *Exp. Fluids*, 17: 405–414.

Arcoumanis C, Whitelaw D S, Whitelaw J H (1996), "Breakup of droplets of newtonian and non–newtonian fluids", *Atomization Sprays*, 6:245–256.

Babinsky E, Sojka PE (2002), "Modeling drop size distributions", *Prog. Energy Combust. Sci.*, 28(4): 303–329.

Batchelor G K (1976), "An Introduction to Fluid Dynamics", *Cambridge Univ. Press*, London.

Batchelor G K (1987), "The stability of a large gas bubble rising through liquid", *J. Fluid Mech.,* 184: 399–422.

Berthoumieu P, Carentz H, Villedieu P, Lavergne G (1999), "Contribution to droplet breakup analysis", *Int. J. Heat Fluid*, 20: 492–498.





Bogoyavlenskiy V A (1999), "Differential criterion of a bubble collapse in viscous liquids", *Phys. Rev.* E, 60(1): 504–508.

Borisov AA, Gelfand BE, Natanzon MS, Kossov OM (1981), "Droplet breakup regimes and criteria for their existence", *J. Engin. Phys.*, 40(1): 44-49.

Bremond C, Villermaux V (2005), "Bursting thin liquid films", *J. Fluid Mech,* 524: 121–130.

Brodkey R S (1967), "Formation of drops and bubbles. In: The phenomena of fluid motions", Addison-Wesley, Reading.

Cao X–K, Sun Z–G, Li W–F, Liu H–F, Yu Z–H (2007), "A new breakup regime of liquid drops identified in a continuous and uniform air jet flow", *Phys. Fluids*, 19(5): 057103.

Chandrasekhar S (1961), "Hydrodynamic and hydromagnetic stability", *Oxford Univ. Press*, London.

Chou W H, Hsiang L P, Faeth G M (1997), "Temporal properties of drop breakup in the shear breakup regime", *Int. J. Multiphase Flow*, 23(4): 651–669.

Chou W H, Faeth G M (1998), "Temporal properties of secondary drop breakup in the bag breakup regime", *Int. J. Multiphase Flow*, 24(6): 889–912.

Clift R, Grace JR, Weber ME (1978), "Bubbles, Drops, and Particles", New York: Academic Press.

Cohen RD (1994), "Effect of viscosity on drop breakup", *Int. J Multiphase Flow*, 20(1): 211–216.





Culick F E C (1960), "Comments on a ruptured soap film", *J. Appl. Phys.*, 31: 1128.

Dai Z, Faeth G M (2001), "Temporal properties of secondary drop breakup in the multimode breakup regime", *Int. J. Multiphase Flow,* 27(2): 217–236.

Davies R M, Taylor G (1950), "The mechanics of large bubbles rising through extended liquids and through liquids in tubes", *Proc. R. Soc. Lond. A*, 200(1062): 375–390.

Debrégeas G, Martin P, Brochard-Wyart F (1995), "Viscous bursting of suspended films", *Phys. Rev. Lett.*, 75(21): 3886 – 3889.

Dodd K N (1960), "On the disintegration of water drops in an airstream", *J. Fluid Mech,* 9(2): 175–182.

Drazin P G, Reid W H (1981), "Hydrodynamic Stability", *Cambridge Univ. Press*.

Eggers J, Villermaux E (2008), "Physics of liquid jets", *Rep. Prog. Phys.,* 71: 036601.

Engel O G (1958), "Fragmentation of water drops in the zone behind an air shock", *J. Res. Natl. Bur.Stand.* 60:245–80.

Faeth G M, Hsiang L–P, Wu P K (1995), "Structure and breakup properties of sprays", *Int. J. Multiphase Flow*, 21: 99–127.

Fakhri A, Rahimian M H (2009), "Simulation of falling droplet by the lattice Boltzmann method", *Commun. Nonlinear Sci. Numer. Simulat.,* 14: 3045–3046.





Flock A K , Guildenbecher  D R, Chen J, Sojka P E, Bauer H–J (2012), " Experimental statistics of droplet trajectory and air flow during aerodynamic fragmentation of liquid drops ", *Int. J. Multiphase Flow*, 47: 37–49.

Flower W D (1928), "The terminal velocity of drops", *Proc. Phys. Soc. London*, 40: 167–176.

Gast L, (1991), *PhD* Thesis, Rutgers, The State University of New Jersey.

Gelfand BE, Gubin SA, Kogarko SM, Komar SP (1973), "Singularities of the breakup of viscous liquid droplets in shock waves", *J. Eng. Phys.*, 25(3): 1140–1142.

Gelfand B E (1996), "Droplet breakup phenomena in flows with velocity lag", *Prog. Energy Combust. Sci.*, 22(3): 201–265.

Gordon G (1959), "Mechanism and speed of breakup of drops", *J. App. Phy.*, 30:1759–1761.

Gorokhovski M, Herrmann M (2008), "Modeling primary atomization", *Annu. Rev. Fluid Mech.*, 40: 343–366.

Guildenbecher D R (2010), "Secondary Atomization of Electrostatically Charged Drops", Ph.D. thesis, Purdue University.

Guildenbecher D R, López-Rivera C, Sojka P E (2009), "Secondary atomization", *Exp. Fluids*, 46: 371–402.

Haas F C (1964), "Stability of droplets suddenly exposed to a high velocity gas stream", *AIChE J.*, 10: 920–924.





Han J, Tryggvason G (1999), "Secondary breakup of axisymmetric liquid drops. I. Acceleration by a constant body force", *Phys. Fluids*, 11(12): 3650–3667.

Han J, Tryggvason G (2001), "Secondary breakup of axisymmetric liquid drops. II. Impulsive acceleration", *Phys, Fluids*, 13(6): 1554–1565.

Hanson A R, Domich E G, Adams H S (1963), " Shock tube investigation of the breakup of drops by air blasts", *Phys. Fluids*, 6: 1070–1080.

Harper E Y, Grube G W, Chang I–D (1972), "On the breakup of accelerating liquid drops", *J. Fluid Mech.*, 52(3): 565–591.

Helenbrook BT, Edwards CF (2002), "Quasi-steady deformation and drag of uncontaminated liquid drops", *Int. J. Multiphase Flow*, 28(10): 1631–1657.

Hinze J O (1955), "Fundamentals of the hydrodynamic mechanism of splitting in dispersion processes", *AIChE J.*, 1(3): 289–295.

Hinch E J (1991), "Perturbation methods", *Cambridge Univ. Press.*

Hirabara H and Kawahashi M (1992), "Experimental investigation of viscous effects upon a breakup of droplets in high-speed air flow", *Exp. Fluids*, 13: 423–428.

Hsiang L–P, Faeth G M (1992), "Near limit drop deformation and breakup", *Int. J. Multiphase Flow*, 18(5): 635–652.

Hsiang L–P, Faeth G M (1993), "Drop properties after secondary breakup", *Int. J. Multiphase Flow*, 19(5): 721–735.





Hsiang L–P, Faeth GM (1995), "Drop deformation and breakup due to shock wave and steady disturbances", *Int. J Multiphase Flow*, 21(4): 545–560.

Hwang S S, Liu Z, Reitz RD (1996), "Breakup mechanisms and drag coefficients of high–speed vaporizing liquid drops", *Atomization Sprays*, 6(3): 353–376.

Igra D, Takayama A (2001), "Investigation of aerodynamic breakup of a cylindrical water droplet", *Atomization Sprays,* 11(2): 167–185.

Jalaal M, Mehravaran K (2012), "Fragmentation of falling liquid droplets in bag breakup mode", *Int. J. Multiphase Flow*, 47: 115–132.

Joseph D D, Belanger J, Beavers GS (1999), "Breakup of a liquid drop suddenly exposed to a high-speed airstream", *Int. J. Multiphase Flow*, 25(6–7): 1263-1303.

Joseph D D, Beavers G S, Funada T (2002), "Rayleigh–Taylor instability of viscoelastic drops at high Weber numbers", *J. Fluid Mech.* 453: 109–132.

Joseph D D (2003), "Rise velocity of a spherical cap bubble", *J. Fluid Mech.*, 488: 213–223.

Joseph D, Funada T, Wang J (2008), "Potential flows of viscous and viscoelastic fluids", *Cambridge Univ. Press.*

Krechetnikov R (2009), "Rayleigh–Taylor and Richtmyer–Meshkov instabilities of flat and curved interfaces", *J. Fluid Mech.*, 625: 387–410.

Krzeczkowski, S A (1980), "Measurement of liquid droplet disintegration mechanism ", *Int. J. Multiphase Flow*, 6: 227–239.





Lamb H (1932), "Hydrodynamics", 6th edition, *Cambridge Univ. Press*.

Landau L D, Lifshitz E M (1989), "Fluid Mechanics", *Pergamon*.

Lane W R, (1951), "Shatter of drops in streams of air", *Ind. Eng. Chem.*, 43: 1312–1317.

Leal L G (1992), "Laminar flow and convective transport processes: Scaling Principles and Asymptotic Analysis", *Butterworth–Heinemann Series in Chemical Engineering*.

Lebedev N (1972), "Special functions and their applications", *Dover Publications Inc.,* New York.

Lee C H, Reitz R D (2000), "An experimental study of the effect of gas density on the distortion and breakup mechanism of drops in high speed gas stream", *Int. J. Multiphase Flow* 26: 229–244.

Lee C S, Reitz R D (1999), "Modeling the effects of gas density on drop trajectory and breakup of high-speed liquid drops", *Atomization Sprays*, 9(5): 4979–511.

Lee C S, Reitz R D (2001), "Effect of liquid properties on the breakup mechanism of high-speed liquid drops", *Atomization Sprays*, 11(1): 1–19.

Lefebvre A H (1989), "Atomization and Sprays", Hemisphere Publishing Corporation, New York.

Lewis DJ (1949), "The instability of liquid surfaces when accelerated in a direction perpendicular to their planes II", *Proc. R. Soc. Lond.* A, 202(1068):81–90.





Lenard P (1904), *Über Regen. Meteorol. Z.,* 6: 249–262.

Lhussier H, Villermaux E (2011), "The destabilization of an initially thick liquid sheet edge", *Phys, Fluids*, 23: 091705(1) – 091705(4).

Lhuissier H, Villermaux E (2012), "Bursting bubble aerosols", *J. Fluid Mech.*, 696: 5–44.

Lin S P (2003), "Breakup of liquid sheets and jets", *Cambridge Univ. Press.*

Liu Z, Reitz R D (1997), "An analysis of the distortion and breakup mechanisms of high speed liquid drops", *Int. J. Multiphase Flow*, 23:631–650.

López-Rivera C (2010), "Secondary breakup of inelastic non-Newtonian liquid drops", Ph.D. thesis, Purdue University.

Marmottant P, Villermaux E (2004), "On spray formation", *J. Fluid Mech.*, 498: 73–111.

Mehrabian H, Feng J J (2013), "Capillary breakup of a liquid torus", *J. Fluid Mech.*, 717: 281-292.

Mc Donald J E (1954), "The shape and aerodynamics of large drops", *J. Meteor.*, 11:478–494.

Ng C–L, Sankarakrishnan R, Sallam K A (2008), "Bag breakup of nonturbulent liquid jets in crossflow", *Int. J. Multiphase Flow*, 34(3): 241–259.

O'Brien V (1961), "Why raindrops break up – Vortex instability", *J. Meteor.*, 18: 549–552.





O'Rourke P J, Amsden A A (1987), "The TAB method for numerical calculations of spray droplet breakup", *SAE Technical Paper*, 87: 2089.

Ortiz C, Joseph D D, Beavers G S (2004), "Acceleration of a liquid drop suddenly exposed to a high-speed airstream", *Int. J. Multiphase Flow*, 30: 217–224.

Padrino J C, Funada T, Joseph D D (2008), "Purely irrotational theories for the viscous effects on the oscillations of drops and bubbles", *Int. J Multiphase Flow*, 34(4): 61–75.

Park J H, Yoon Y, Hwang SS (2002), "Improved TAB model for prediction of spray droplet deformation and breakup", *Atomization Sprays*, 12(4): 387–401.

Park S P, Kim S, Lee C S (2006), "Breakup and atomization characteristics of monodispersed diesel droplets in a cross-flow air stream", *Int. J. Multiphase Flow*, 32: 8079–826.

"Physical properties of glycerine and its solutions" (1963), *Glycerine Producers Association*.

Pilch M, Erdman CA (1987), "Use of breakup time data and velocity history data to predict the maximum size of stable fragments for acceleration-induced breakup of a liquid drop", *Int. J. Multiphase Flow*, 13(6): 741–757.

Plesset M S, Prosperetti A (1977), "Bubble Dynamics and Cavitation", *Ann. Rev. Fluid Mech.,* 9: 145–85.

Ranger AA, Nicholls JA (1969), "The Aerodynamic Shattering of Liquid Drops", *AIAA J.*, 7(2): 285–290.





Reitz R D (1987), "Modeling atomization processes in high-pressure vaporizing sprays", *Atomisation Spray Technol.,* 3: 307–337.

Reyssat É, Chevy F, Biance A–L, Petitjean L, Quéré D (2007), "Shape and instability of free-falling liquid globules", *Europhysical Letters*, 80: 34005p1–34005p5.

Rodriguez M, Manuel G, Muiioz M J, Martin R (1994), "Viscosity of Triglycerides + Alcohols from 278 to 313 K", *J. Chem. Eng. Data*, 39: 102–105.

Scharfman B E, Techet A H (2012), "Bag instabilities", *Phys. Fluids*, 24(9): 091112.

Schmelz F, Walzel P (2003), "Breakup of liquid droplets in accelerated gas flows", *Atomization Sprays*, 13(4): 357–372.

Scriven L E (1960), "Dynamics of a fluid interface equation of motion for Newtonian surface fluids", *Chem. Eng. Sci.*, 12(2): 98-108.

Shraiber A A, Podvysotsky A M, Dubrovsky V V (1996), "Deformation and breakup of drops by aerodynamic forces", *Atomization and Sprays*, 6: 667–692.

Simpkins P G (1971), "On the distortion and breakup of suddenly accelerated droplets", *AIAA J.*, 71–325.

Simpkins P G, Bales E L (1972), "Water-drop response to sudden accelerations", *J. Fluid. Mech.,* 55: 629–639.

Snyder S E (2011), "Secondary Atomization of Elastic Non-Newtonian Liquid Drops", MSME thesis, Purdue University.





Spilhaus A F (1948), "Raindrop size, shape, and falling speed", *J. Meteor.*, 5:108–110.

Strogatz S H (1994), "Nonlinear Dynamics and Chaos: With Applications to Physics, Biology, Chemistry, And Engineering", *Perseus Books Publishing*.

Tarnogrodzki A (1993), "Theoretical prediction of the critical Weber number", *Int. J. Multiphase Flow*, 19 (3): 329–336.

Tarnogrodzki A (2001), "Theory of free fall breakup of large drops", *Int. J. Mech. Sci.*, 43 (4): 883–893.

Taylor G (1950), "The instability of liquid surfaces when accelerated in a direction perpendicular to their planes. I", *Proc. R. Soc. London Ser. A*, 201(1065): 192–196.

Taylor G (1959), "The dynamics of thin sheets of fluid I: Water bells", *Proc. R. Soc. London Ser. A*, 253(1274): 289–295.

Taylor G (1959), "The dynamics of thin sheets of fluid II: Waves on fluid sheets", *Proc. R. Soc. Lond. A*, 253(1274): 296–312.

Taylor G (1959), "The dynamics of thin sheets of fluid III: Disintegration of fluid sheets", *Proc. R. Soc. Lond. A*, 253(1274): 313–321.

Theofanous T G, Li G J, Dinh T N (2004), "Aerobreakup in rarefied supersonic gas flows", *J. Fluids Eng.*, 126(4): 516–527.

Theofanous, T G, Li, G J (2008), "On the physics of aerobreakup", *Phy. Fluids*, 20: 052103.





Theofanous T G (2011), "Aerobreakup of Newtonian and viscoelastic drops", *Annu. Rev. Fluid Mech.,* 43:661–90.

Thompson J J, Newall H F (1885), "On the formation of vortex rings by drops falling in liquid and some allied phenomena", *Proc. R . Soc. London,* 39:418.

Tomotika S (1935), "On the instability of a cylindrical thread of a viscous liquid surrounded by another viscous fluid", *Proc. R. Soc. Lond. A*, 150: 322–337.

Tomotika S (1936), "Breaking up of a drop of viscous liquid immersed in another viscous fluid which is extending at a uniform rate", *Proc. R. Soc. Lond.* A, 153(879): 302–318.

Villermaux E (2007), "Fragmentation", *Annu. Rev. Fluid Mech.,* 39: 419–446.

Villermaux E, Bossa B (2009), "Single-drop fragmentation determines size distribution of raindrops", *Nat. Phys.*, 5: 697–702.

Villermaux E, Bossa B (2011), "Drop fragmentation on impact", *J. Fluid Mech.*, 668: 412–435.

Wert KL (1995), "A rationally-based correlation of mean fragment size for drop secondary breakup", *Int. J. Multiphase Flow*, 21(6): 1063-1071.

Wierzba A, Takayama K (1988), "Experimental investigation of the aerodynamic breakup of liquid drops", *AIAA J.*, 26(11): 1329–1335.

Wierzba A (1990), "Deformation and breakup of liquid drops in a gas stream at nearly critical Weber numbers", *Exp. Fluids*, 9: 59–64.





Wu M, Cubaud T, Ho C–M (2004), "Scaling law in liquid drop coalescence driven by surface tension", *Phys. Fluids*, 16(7): L51–L54.

Wu P K, Hsiang L P, Faeth G M (1995), "Aerodynamic effects on primary and secondary breakup", *Prog. Astronaut. Aeronaut.*, 169: 247.

Wu Z–N (2003), "Approximate critical Weber number for the breakup of an expanding torus", *Acta Mech.*, 166: 231–239.

Zhao H, Liu H–F, Cao X–K, Li W–F, Xu J–L (2011), "Breakup characteristics of liquid drops in bag regime by a continuous and uniform air jet flow", *Int. J. Multiphase Flow*, 37 (3): 530–534.

Zhao H, Liu H–F, Cao X–K, Li W–F (2011), "Experimental Study of Drop Size Distribution in the Bag Breakup Regime", *Ind. Eng. Chem. Res.*, 50: 9767–9773.

Zhao H, Liu H–F, Xu J–L, Li W–F (2011), "Secondary breakup of coal water slurry drops", *Phys. Fluids*, 23: 113101.

Zhao H, Liu H-F, Xu J-L, Li W-F, Lin K-F (2013), "Temporal properties of secondary drop breakup in the bag -stamen breakup regime", *Phy. Fluids*, 25: 054102.

Zhao H, Liu H–F, Li W–F, Xu J–L (2010), "Morphological classification of low viscosity drop bag breakup in a continuous air jet stream", *Phys. Fluids*, 22: 114103.


APPENDIX



APPENDIX

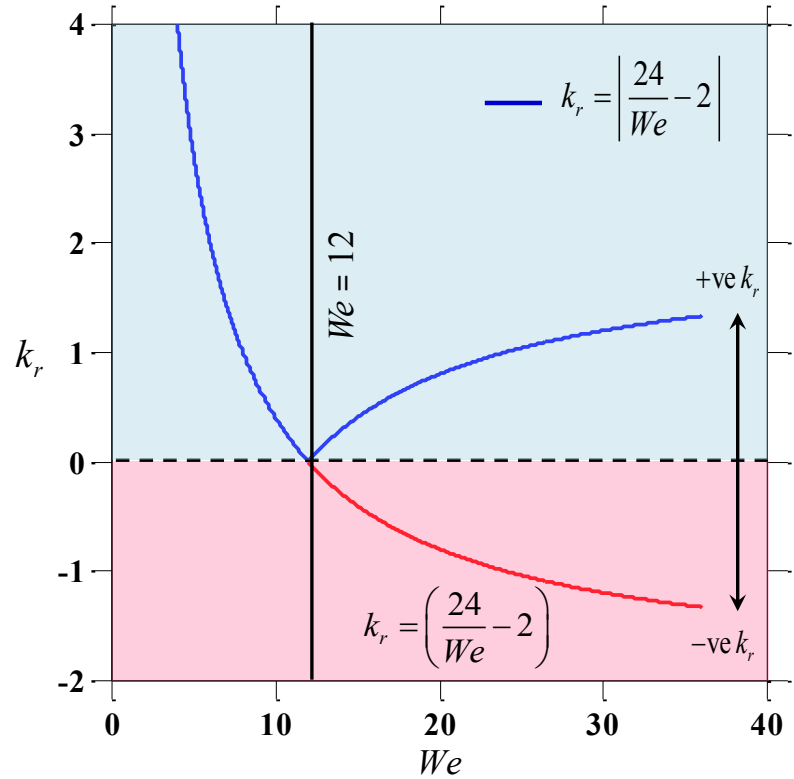

Figure 1: Variation of $k_r$ with $We$

The above figure shows that $k_r$ to remain positive must belong to the blue region of the plot. Once $We_{c\,Oh \to 0} = 12$ is reached $k_r$ flips sign to ensure that the restoring force remains positive.

VITA



VITA

Varun Kulkarni was born on February 14$^{th}$, 1985 in Mumbai, India. Upon completion of his high school education he attended PES Institute of Technology, Bangalore, India for his undergraduate studies where he pursued Mechanical Engineering. Following his graduation he joined the Department of Aerospace Engineering, Indian Institute of Science, Bangalore for his Masters. Thereafter, he worked at 3DPLM, Pune, India for a year before pursuing his PhD at School of Mechanical Engineering at the Purdue University, West Lafayette. Currently, he is looking for a Post-Doctoral position.